\documentclass[12pt,letterpaper]{article}  
\pdfoutput=1

\usepackage[includeheadfoot,
            marginratio={1:1,2:3}, 
            width=412pt, 
            height=688pt,]{geometry}

\usepackage{amsmath}
\usepackage{amsfonts}
\usepackage{amssymb}
\usepackage{mathtools}
\usepackage{graphicx}
\usepackage{booktabs}
\usepackage{adjustbox}
\usepackage[utf8]{inputenc}
 \usepackage[splitrule]{footmisc} %% The splitrule option draws a full width rule above the continued part of the footnote as a visual cue to readers.

\usepackage{empheq}
\usepackage{paralist}
\usepackage{cite}
\usepackage[hidelinks]{hyperref}
\usepackage{xcolor}
\usepackage{tensor}

\newcommand{\nc}{\newcommand}
\nc{\lb}{\llbracket}
\nc{\rb}{\rrbracket}
\nc{\gl}{\llbracket}
\nc{\gr}{\rrbracket}
\nc{\del}{\partial}
\nc{\tri}{\hspace{-3.5pt}\vartriangle\hspace{-3.5pt}}
\nc{\blacktri}{\blacktriangle}

\nc{\eq}[1]{\begin{equation}
                     \begin{split} #1 \end{split}
                     \end{equation}}
\nc{\ul}{\underline}
\nc{\ov}{\overline}

\nc{\fa}{\hat}
\nc{\fb}{\MakeUppercase}
\nc{\fc}{\tilde }
\nc{\Lie}{{\cal L}} 
\nc{\lambdabar}{{\mkern0.75mu\mathchar '26\mkern -9.75mu\lambda}}

\allowdisplaybreaks[2]
\numberwithin{equation}{section}

\DeclareUnicodeCharacter{2212}{-}
\begin{document}

\vspace*{-1.5cm}
\begin{flushright}
  {\small
  MPP-2020-213\\
  }
\end{flushright}

\vspace{1.5cm}
\begin{center}
{\LARGE
De Sitter  Quantum Breaking,  \\[0.2cm]
Swampland Conjectures and Thermal Strings \\[0.2cm]
} 
\vspace{0.4cm}

\end{center}

\vspace{0.35cm}
\begin{center}
 Ralph Blumenhagen$^{1}$,
Christian Knei\ss l$^{1,2}$,
Andriana Makridou$^{1}$ 
\end{center}

\vspace{0.1cm}
\begin{center} 
\emph{
$^{1}$ 
Max-Planck-Institut f\"ur Physik (Werner-Heisenberg-Institut), \\ 
   F\"ohringer Ring 6,  80805 M\"unchen, Germany } 
   \\[0.1cm] 
\vspace{0.25cm} 
\emph{$^{2}$ Ludwig-Maximilians-Universit{\"a}t M\"unchen, Fakult{\"a}t f{\"u}r Physik,\\ 
Theresienstr.~37, 80333 M\"unchen, Germany}\\
\vspace{0.2cm}

\vspace{0.3cm} 
\end{center} 

\vspace{0.5cm}

%%%%%%%%%%%%%%%%%%%%%%%%%%%%%%%%%%%%%%%%%%%%%%%
%%%%%%%%%%%%%%%%%%%%%%%%%%%%%%%%%%%%%%%%%%%%%%%
%%%%%%%%%%%%%%%%%%%%%%%%%%%%%%%%%%%%%%%%%%%%%%%
%%%%%%%%%%%%%%%%%%%%%%%%%%%%%%%%%%%%%%%%%%%%%%%
%%%%%%%%%%%%%%%%%%%%%%%%%%%%%%%%%%%%%%%%%%%%%%%
%%%%%%%%%%%%%%%%%%%%%%%%%%%%%%%%%%%%%%%%%%%%%%%
%%%%%%%%%%%%%%%%%%%%%%%%%%%%%%%%%%%%%%%%%%%%%%%
%%%%%%%%%%%%%%%%%%%%%%%%%%%%%%%%%%%%%%%%%%%%%%%

\begin{abstract}
We argue that under certain assumptions the quantum break time approach and the
trans-Planckian censorship  conjecture both  lead to de Sitter swampland constraints
of the same functional form.
It is a well known fact that the quantum
energy-momentum tensor in the Bunch-Davies vacuum
computed in the static patch of dS breaks some of the isometries. 
Proposing   that this 
is a manifestation of quantum breaking of dS, we analyze
some of its  consequences. 
In particular, this leads to  a thermal matter component that can 
be generalized to string theory in an obvious way.
Imposing a censorship of  quantum breaking, 
we recover the no eternal inflation bound  in the low temperature
regime, while  the stronger bound from the dS swampland conjecture 
follows under a few reasonable assumptions about the still
mysterious, presumably topological, high-temperature regime of string
theory. 
\end{abstract}

\clearpage

\tableofcontents

%%%%%%%%%%%%%%%%%%%%%%%%%%%%%%%%%%%%%%%%%%%%%%%
%%%%%%%%%%%%%%%%%%%%%%%%%%%%%%%%%%%%%%%%%%%%%%%
%%%%%%%%%%%%%%%%%%%%%%%%%%%%%%%%%%%%%%%%%%%%%%%
%%%%%%%%%%%%%%%%%%%%%%%%%%%%%%%%%%%%%%%%%%%%%%%
%%%%%%%%%%%%%%%%%%%%%%%%%%%%%%%%%%%%%%%%%%%%%%%
%%%%%%%%%%%%%%%%%%%%%%%%%%%%%%%%%%%%%%%%%%%%%%%
%%%%%%%%%%%%%%%%%%%%%%%%%%%%%%%%%%%%%%%%%%%%%%%
%%%%%%%%%%%%%%%%%%%%%%%%%%%%%%%%%%%%%%%%%%%%%%%

%\newpage

\section{Introduction}
\label{sec:intro}

The de Sitter (dS) space-time is one of the most studied solutions of Einstein's equations, partially because of  its value for cosmological considerations. Its possible instability has been a matter of great interest already since the 80s \cite{Mottola:1984ar, Mottola:1985qt} as it could provide an explanation for the cosmological constant problem. 
By now there exist indications that quantum gravity does
not admit long-lived de Sitter solutions. The most developed
candidate for a theory of quantum gravity, namely string theory, does not yet feature a fully controlled meta-stable dS minimum that is beyond
any doubt. The KKLT construction \cite{Kachru:2003aw}, which gives a
dS minimum as the result of a non-trivial three step procedure, still
lies in the center of an on-going debate regarding its
viability.
Several lines of reasoning can lead to conjectures and bounds which qualitatively forbid
long-lived dS vacua, yet their mutual interrelations and respective regimes
of validity are not clear. The aim of
the present paper is to put some of these conjectures into
perspective, and to do so, let us start by briefly reviewing the relevant
bounds and their  origins.

\subsubsection*{Swampland bounds on the scalar potential}

In the context of the string swampland, put forward in
\cite{Vafa:2005ui} and more recently reviewed in \cite{Palti:2019pca,vanBeest:2021lhn,Grana:2021zvf}, 
the difficulty in  constructing dS solutions  with a bona fide ten-dimensional uplift is thought to be no coincidence.  In particular, 
it points towards the idea of de Sitter not being allowed in string
theory at all  \cite{Brennan:2017rbf,Danielsson:2018ztv}.
In \cite{Obied:2018sgi} the  quantitative dS swampland conjecture was proposed,
according to which the effective four-dimensional scalar potential $V$
always satisfies 
\eq{
\label{dswamp}
                 M_{\rm pl}  |V'|\ge c \cdot V\, ,
}
where $c$ is an ${O}(1)$ positive number and in particular
independent of $V$, though possibly $n$-dependent when considering $n$ space-time dimensions.
In \cite{Agrawal:2018own} the observational consequences of such a
bound were discussed.

Clearly, this conjecture is trivially satisfied for extrema of the potential whenever $V<0$, while it forbids dS vacua with $V>0$.
It is worth noting that  for $V>0$ the dS swampland bound can
alternatively be expressed as a lower bound for the value of the
slow-roll parameter $\epsilon$. Such a bound, in a much more
restricted setup, was already computed in \cite{Hertzberg:2007wc}
using a simple scaling argument, and was later generalized in \cite{Obied:2018sgi}.

Soon it became clear that the conjecture should be slightly relaxed to allow for certain dS maxima/saddle points \cite{Denef:2018etk, Andriot:2018wzk}. 
The refined dS conjecture \cite{Ooguri:2018wrx-2} supplemented the initial conjecture with the alternative clause: 
\eq{
\label{rdswamp}
M_{\rm pl}^2   |V''|\ge  c' \cdot V\, ,
}
where $c'$ is once again an ${O}(1)$ constant.
The same generalization of the dS conjecture was already proposed in
\cite{Garg:2018reu} in terms of the two slow-roll parameters
$\epsilon, \eta$.

While the dS swampland conjecture is mostly motivated by string tree-level examples, a conjecture
with a  more direct physical motivation was proposed in \cite{Bedroya:2019snp}.
This is the trans-Planckian censorship conjecture (TCC), which 
states that sub-Planckian fluctuations in an expanding quasi dS space should not become classical. 
Applying this conjecture to scalar fields and focusing only on
positive potentials, one  gets  bounds very similar to  the dS
conjecture. In particular, in the asymptotic limit  of the field space
TCC assumes the form 
\eq{
\label{tcc}
M_{\rm pl}^{\frac{n-2}{2}} \left( \frac{|V'|}{V}\right)_\infty \ge \frac{2}{\sqrt{(n-1)(n-2)}}\, .
}
This TCC limiting case is of the same form as the dS conjecture, while also providing an explicit prediction for the value of $c$.
If we do not invoke this asymptotic limit, the TCC is in general weaker than the dS conjecture, in the sense that sufficiently short-lived dS vacua are permitted.
The physical motivation of the TCC has been debated \cite{Dvali:2020cgt, Burgess:2020nec}  recently. Here we do not intend to enter this discussion, but throughout this paper we rather consider TCC as a working assumption from which other swampland constraints can be inferred.

Another proposal which restricts the scalar potential and its
derivatives is the ``no eternal inflation''
principle.  In  \cite{Rudelius:2019cfh} necessary conditions for (no)
eternal inflation were inferred by solving the Fokker-Planck equation
for stochastic inflation. For several types of potentials, the no eternal inflation
bounds exhibited remarkable similarities to several formulations of
the (refined) dS conjecture. This hinted at a no eternal inflation principle as a 
possible deeper reason behind the swampland conjectures.  In the present paper we will mostly be concerned with slowly-rolling scalar fields, so we start with the case of a four-dimensional  linear potential. There, the no eternal inflation principle
imposes the bound
\eq{
\label{noeternalinfl}
M_{\rm pl}\frac{|V'|}{V}> \frac{\sqrt {2}}{2 \pi }\bigg( \frac{V}{M_{\rm pl}^4}\bigg)^{1/2} \, .
}
This is remarkably similar to the dS conjecture \eqref{dswamp}, but
now the right-hand side is still  $V$-dependent, relaxing the bound significantly for small and positive $V$. 
Let us note that  conditions necessary for eternal inflation, possibly
differing in the ${O}(1)$ constant,
were already obtained in  \cite{Linde:2005ht, Page:1997vc,Creminelli:2008es}.
The compatibility of eternal inflation with the dS swampland conjecture has also been studied in \cite{Matsui:2018bsy}.

Moreover, the bound \eqref{noeternalinfl} was generalized to $n$ dimensions in \cite{Bedroya:2020rac}, where it assumes the form
\eq{
\label{noeternalinflndim}
M_{\rm pl}^{\frac{n-2}{2}} \frac{|V'|}{V}> c'' \cdot \bigg( \frac{V}{M_{\rm pl}^n}\bigg)^{\frac{n-2}{4}}, 
}
with $c''$ again an ${O}(1)$  positive constant. 
In the following, we will refer to any bound of the form $|V'|/V> {c}\cdot V^{\frac{n-2}{4}}$ as a no eternal inflation bound, irrespective of the precise value of this positive constant  ${c}$. Moreover, we are setting the reduced Planck mass to one, unless explicitly stated.

One should keep in mind that different types of potentials result in different bounds.
For a four-dimensional quadratic hilltop potential, 
the no eternal inflation principle translates to the bound 
\eq{
\label{noeternalinfhilltop}
M_{\rm pl}^2\frac{|V''|}{V}>3\,.
}
Provided $c'>3$, this is  compatible with the refined dS conjecture \eqref{rdswamp}. 
Similar bounds, up to ${O}(1)$ factors, for eternal hilltop inflation were also derived in \cite{Barenboim:2016mmw, Kinney:2018kew}.

An intricate relation
between the TCC and the no eternal inflation principle was recently
revealed in \cite{Bedroya:2020rac}. There, a sequence
of short-lived dS bubbles, transitioning from one to the next through
non-perturbative membrane nucleation, was described in terms of a dual
low-energy effective scalar potential. 
Imposing TCC was sufficient to find the allowed region in the
parameter space, which turned out to 
marginally exclude eternal
inflation, thus leading to a  bound  of the form $|V'|> c\cdot V^{3/2}$.

\subsubsection*{Quantum breaking of de Sitter and relation to swampland}

The idea of having an upper bound on the life-time of de Sitter is not new.
In a different line of research \cite{Dvali:2013eja,Dvali:2014gua,Dvali:2017eba},
a corpuscular picture of de Sitter space was proposed, which allowed to
capture quantum effects invisible to the usual semiclassical treatment of gravity.
Those effects are found to induce a finite life-time for classical dS 
solutions, the so-called quantum break time.
More specifically, the dS solution is viewed as a coherent state
of  gravitons over Minkowski space. 
Decoherence of this  state, i.e. quantum scattering of gravitons from
the coherent state, leads then to the quantum break time,
that is the finite time-scale after which the mean field description ceases being valid.
A general expression for this life-time was given by
$t_Q\sim t_{\rm cl}/\alpha$, where $ t_{\rm cl}$ denotes the
characteristic time scale of the system and $\alpha$ the quantum
interaction strength of the constituents. For a four-dimensional 
inflating phase in pure Einstein gravity,  the first parameter is the Hubble time  $t_{\rm cl}=H^{-1}$
and the second one $\alpha=H^2/M_{\rm pl}^2$, i.e. an effective strength
of graviton scattering for the characteristic momentum transfer $H$. Thus,  the quantum break
time becomes $t_q\sim M_{\rm pl}^2/H^3$\footnote{In \cite{Aalsma:2019rpt} the same time scale was found for the so-called Unruh-de Sitter state.}.

In view of the stringy swampland conjectures, 
this picture was extended in \cite{Dvali:2018fqu, Dvali:2018jhn} by suggesting that quantum breaking should not occur. The theory must censor it by providing a (classical) mechanism that leads to a faster decay of de Sitter.
Such a behavior is for instance exhibited by a sufficiently fast rolling scalar field with
slow-roll parameter $\epsilon$ and the associated classical time-scale 
$ t_{\rm cl}\sim 1/(\epsilon H)$.  Requiring $ t_{\rm cl}\lesssim t_q$ leads to the following bound on the gradient of the potential
\eq{
\label{qBbound}
     M_{\rm pl}\frac{|V'|}{V}\gtrsim  c \cdot \bigg( {\frac{V}{  M^4_{\rm pl}}} \bigg)^{\frac{1}{2}},\
    } 
which takes the same form as the no eternal inflation bound
and is therefore weaker than the dS swampland conjecture.
The only way to get the stronger bound of the dS conjecture 
would be a scaling $t_q\sim H^{-1}$.

The present work examines whether also the dS swampland
bound \eqref{dswamp} can be derived in a similar fashion  as a quantum
breaking bound. Since
the bound \eqref{qBbound} is derived using a quantum description of
classical gravity, it is natural to suspect that the answer is related
to a generalization to a more complete  quantum gravitational theory like
string theory, loop quantum gravity or any of the possible candidates.
It has already been pointed
out in  \cite{Dvali:2018fqu, Dvali:2018jhn}  that in perturbative
string theory  the bound might get stronger. 
One proposal was that in string theory the natural estimate for the
  interaction strength is the string coupling constant
  $\alpha=g_s^2$. This  leads indeed to the required 
  $t_Q\sim H^{-1}$  scaling  but would also imply a factor of $g_s^2$
  in the dS swampland bound \eqref{dswamp}.  Notice that  such a factor is
  not present  in the  TCC bound \eqref{tcc}.

It would be interesting to generalize the approach
\cite{Dvali:2013eja,Dvali:2014gua,Dvali:2017eba}
to string theory with its additional ingredients like  higher-form
fields and massive excitations\footnote{{In fact, a construction of four-dimensional de Sitter as a
coherent state in full string theory has been proposed in \cite{Brahma:2020htg,Brahma:2020tak}, with a life time consistent with the TCC bound \cite{Bernardo:2020lar}. }}. 
Here, we will follow a different
route that we find easier to generalize to string theory. 
Recall that in the corpuscular picture of \cite{Dvali:2017eba}, the decoherence
is caused by the scattering of gravitons that comprise the coherent state and thus
can be considered  as the backreaction of the quantum state on the classical
geometry.

This is reminiscent of what was discussed in the past \cite{Davies:1977ze,Davies:unknown,Bunch:1978gb}, when in a
semiclassical approach people
computed the expectation value of the energy-momentum tensor
$\langle T_{\mu\nu} \rangle_{\rm BD}$ 
of a scalar field propagating on a curved classical dS background geometry.
Using Friedmann-Lema\^itre-Robertson Walker (FLRW) coordinates
one could define the thermal Bunch-Davies vacuum that led
to an energy-momentum tensor that preserved all the dS isometries.
Considering the backreaction of this quantum contribution on the right-hand side of the Einstein equation only gives rise to a small
redefinition of the cosmological constant. On the other hand,
it is known that performing the same computation in the static
dS patch leads to a qualitatively different result. In this case
$\langle T_{\mu\nu} \rangle_{\rm BD}$ diverges on the dS horizon and
does not preserve the isometries
but instead contains a (matter-like) thermal component with an equation of
state parameter $\omega\ne -1$. For instance, for the 4D conformal case,
one finds  a thermal,
time-independent energy-momentum tensor with an equation of state
$p=\rho/3$ and a temperature equal to the Gibbons-Hawking temperature $T=H/(2\pi)$
of the dS horizon \cite{Gibbons:1977mu}.

If one  includes this contribution in the Einstein equation, it will
cause a deviation from the initial  dS geometry. We will see that the
relevant time scale for this change is the same as the quantum
breaking time of \cite{Dvali:2017eba}. The puzzle of the
prediction of qualitatively different evolutions of dS in FLRW and the
static patch is resolved by censoring quantum breaking, i.e.
by conjecturing that the full theory of quantum gravity even
classically
will not admit eternal dS solutions but at best quasi dS ones,
where e.g. the scalar field is still evolving like in quintessence
models.

Thus, in this paper we will work under the assumption that though
diverging at the horizon, the quantum energy-momentum tensor
in the static patch gives a physically reasonable quantity, at least in
the vicinity of the center of the patch, whose backreaction causes
quantum breaking\footnote{Note that this is very similar in spirit and in fact motivated by the approach of
T. Markannen \cite{Markkanen:2016jhg,
  Markkanen:2016vrp, Markkanen:2017abw}, which however differs
from our approach in that he claims that also the
energy-momentum tensor in FLRW  coordinates does not preserve
the dS isometries.}. In the following we will call this approach the
``quantum backreaction approach''.

\subsubsection*{A pictorial overview and outline}

\begin{figure}[htb]
  \centering
   \includegraphics[width=14cm]{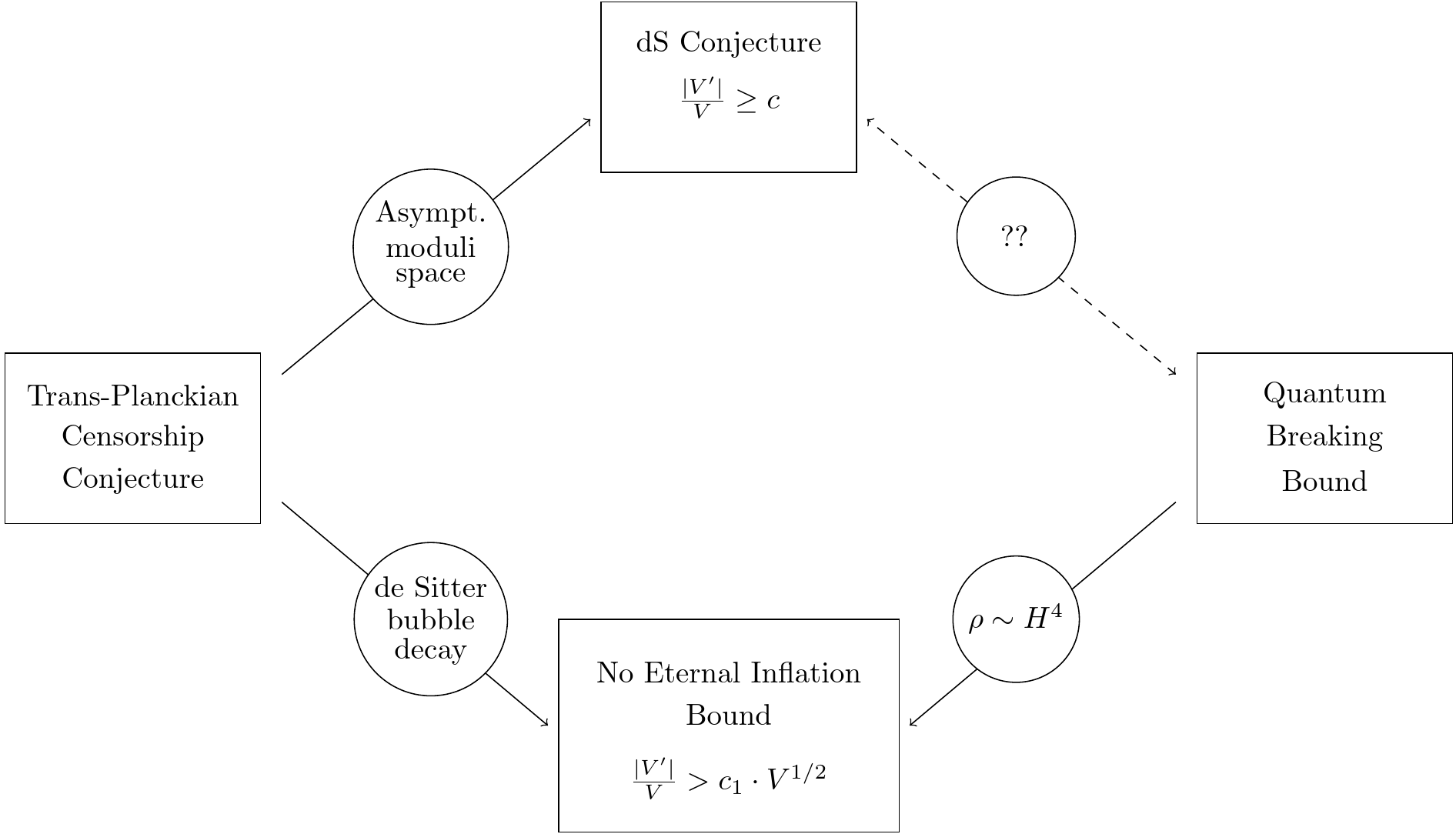}
  \caption{Schematic depiction of bounds for $|V'|/V$ for a  (positive)  scalar potential $V$ in 4D and their origin in terms of swampland conjectures or avoidance of the quantum breaking. $c, c_1$ are positive ${O}(1)$ constants.}
  \label{fig:overview}
\end{figure}

Let us summarize what we have mentioned up to now in a pictorial way in figure \ref{fig:overview}.
There we present only the bounds on $|V'|/V$  for a four-dimensional potential, but 
during the course of the paper we will also consider  the bounds on $|V''|/V$.

We see that as far as the swampland program is concerned, de Sitter
vacua are short-lived if at all allowed.
It is by now well known that the swampland conjectures \cite{Vafa:2005ui,ArkaniHamed:2006dz,Ooguri:2006in,Ooguri:2016pdq, Obied:2018sgi,Ooguri:2018wrx-2, Klaewer:2018yxi, Lust:2019zwm} are not independent but form
a tightly-knit web, with many connections between its nodes\footnote{Several of these interrelations came to light following thermodynamical or entropy related arguments, such as in \cite{Ooguri:2018wrx-2,Seo:2019mfk,Seo:2019wsh,Geng:2019zsx,Bonnefoy:2019nzv,Luben:2020wix}.}. TCC is postulated to be one of the 
central nodes \cite{Bedroya:2020rmd}, and in this paper we ``zoom in"
on only two of its
derivative conjectures. This can be seen on the left-hand side of the
figure, as TCC can lead to
both the dS swampland conjecture
and the no eternal inflation bound. The former case arises when
considering an asymptotic limit in the field space, while the latter
bound is reached through the more complicated dS bubble cascade
decay. On the right-hand side of the figure we have the quantum breaking
approach. Requiring that dS decays fast enough to avoid reaching the quantum break time,  
via both the usual quantum corpuscular
description and the quantum backreaction approach,  leads to the
less strict $|V'|/V>c_1 \cdot V^{1/2}$ no eternal inflation
bound. 

Clearly in this picture there is one link missing that would connect
the censoring of quantum breaking directly to the dS conjecture. 
As already suggested in \cite{Dvali:2018fqu, Dvali:2018jhn} this
link should be related to inherently  quantum gravitational effects.
Our analysis will eventually lead  us to the very intriguing suspicion that it is the high temperature
regime of quantum gravity that provides the missing link. 
There is evidence not only from string theory, but also from other
approaches to quantum gravity\footnote{We
  are indebted to Marco Scalisi for pointing this out to us.} (see
\cite{Carlip:2009kf,Carlip:2017eud}) 
that in this regime the number of degrees of
freedom is  reduced and effectively the theory becomes two-dimensional.
A full description of this phase is yet to be provided, but there are
indications that it might be related to a topological gravity
theory. Note that this proposal is also the basis for the recent idea of
Agrawal, Gukov, Obied, Vafa  \cite{Agrawal:2020xek} that the early
history of the universe is not described by inflation but by such 
a topological gravity theory \cite{Witten:1988xi}. 
  
In the present paper we intend to put more flesh on this idea and
make it concrete in the framework of the quantum backreaction approach. Since in the
course of our arguments we will employ various conceptually different
techniques, we have made  an attempt to provide a self-contained
presentation of the relevant material.   

The paper  is organized as follows:
In section \ref{sec_2} we explain the quantum backreaction  approach
to  quantum breaking in some detail by first reviewing known results
from the literature and recalling  that  some of them are already fixed
by the conformal anomaly. Then we generalize the results for
the energy-momentum tensor in the static patch to 
 higher dimensions and non-conformal scalars.
As we will show, this includes a time-independent,  thermal matter component, which
gives rise to an evolution of the Hubble parameter, leading to a
finite  quantum break time. Moreover, by discussing a non-conformal,
massive scalar in 4D as an enlightening example, we will see that
to a good approximation the energy-density  is given by the familiar  flat space
result.  The only deviation comes from a kind of a resonant behavior
when the Compton wavelength of the massive scalar is equal to the size
of the dS space.
The similarity to the just mentioned flat
space thermal contribution will be 
directly generalized to string theory, thus connecting
to well known approaches to compute thermal one-loop partition functions
in string theory.

In section \ref{sec_3} we recall  the thermodynamics of
strings at finite temperature, focusing on the computation of the
thermal partition function and the related free energy, energy density and
pressure. The salient  new feature   appearing in string theory
is the existence  of the so-called  Hagedorn
temperature $T_H$, where due to the condensation of a stringy winding mode, 
the system is supposed to experience a first order
phase transition. Not much is known about  this new phase of
strings, but the seminal work of Atick-Witten \cite{Atick:1988si} suggested that 
the temperature dependence of the energy density in any dimension is quadratic, 
hence indicating a radical reduction of the number of degrees of freedom.
Such a reduction was previously  also observed in high energy scattering
of strings \cite{Gross:1988ue}.

In section \ref{sec_4} we  study the
behavior of the aforementioned quantities in the low and the high
temperature regimes. In the former,  methods from 
perturbative
string theory are under control whereas for the latter, we have to rely on (naive)
extrapolations and well-motivated guesswork.
 Estimating the quantum break time for these two distinct phases,   we
find that  the low temperature regime leads to a bound on $|V'|/V$ of  the
no eternal inflation principle type, 
while the high temperature regime produces the dS 
swampland conjecture bound.  Therefore, from this perspective there
seems to be a relationship  between the dS swampland conjecture
and the still mysterious high temperature phase of string theory.

\section{Quantum breaking of dS from backreaction}
\label{sec_2}

We start this section by reviewing the computation of the (BD) vacuum
expectation value of  the energy-momentum
tensor for a quantized conformal scalar field in a classical  dS space-time.
We highlight the difference between the results in FLRW coordinates
and in the static patch. Then we provide a systematic approach to
compute the latter in the center of the static patch also for
the non-conformal case. All this will lead us to the conjecture that 
quantum backreaction is a manifestation of quantum breaking.
We will also comment on the similarity to  Rindler space which,
if taken seriously,  will lead to the conjecture that eternal flat
Minkowski space is not a solution of (non-supersymmetric) quantum gravity,  either.

\subsection{Brief review of the conformal case}
\label{sec_Markkanen}

To make our point, let us first consider the simplest  case of a conformally
coupled scalar field $\Phi$ in $n$-dimensional de Sitter space.
For dS space we consider two different coordinate systems. 
First, we have the  Friedmann-Lema\^itre-Robertson-Walker (FLRW) coordinates,
whose line element is given by
\eq{ 
\label{FLRW}
ds^2 =- dt^2 + a^2(t) d\boldsymbol{x}^2 = - dt^2 + e^{2Ht} d\boldsymbol{x}^2\,,
}
with $t \in [-\infty,\infty]$ and the spatial coordinates $x^i \in  [-\infty,\infty]$.
%where $\boldsymbol{x}$ denotes the spatial coordinates.
Introducing the conformal time $\eta$ via $d\eta=dt/a(t)$, this can be expressed as
\eq{ 
\label{FLRWconf}
ds^2 ={\frac{1} {H^2 \eta^2}} \left(- d\eta^2 +  d\boldsymbol{x}^2\right)
}
which makes it evident that the metric is conformally flat.
The FLRW coordinate system is the most common choice for quantizing a
scalar field in de Sitter space.  There exists a family of vacua
extending over the whole FLRW patch which respect the isometries of
dS, the so-called $\alpha$-vacua \cite{Mottola:1984ar,Allen:1985ux}.   A special case among them is the
famous Chernikov-Tagirov or Bunch-Davies (BD) vacuum
\cite{Chernikov:1968zm, Bunch:1978yq}, which is thermal
and leads to a scale-invariant power spectrum consistent with CMB fluctuations.
For further information on dS coordinate systems and the BD vacuum we refer to some standard
references \cite{Birrell:1982ix, Spradlin:2001pw,Parker:2009uva}.

The de Sitter horizon is not explicitly visible in the FLRW patch.
To make it manifest
it is more appropriate to use the metric in the static coordinate system
\eq{
\label{static_de_Sitter}
ds^2 = −(1−H^2r^2)\,d\tau^2+\frac{dr^2}{1−H^2r^2}+r^2d\Omega^{2}_{n-2}\,.   
}
Here the metric components are time-independent and there is
a  coordinate singularity at the horizon $r_h = H^{-1}$. 
While the FLRW patch covers half of the dS manifold,  including a region beyond the horizon,
this does not hold for the static case. 
One has to distinguish the regions inside $0\le r\le H^{-1}$ (region A) and 
outside $r\ge  H^{-1}$  of the horizon (region B), where for the latter case  
$\tau$ becomes spacelike and $r$ timelike (and the metric is no longer strictly speaking ``static").   
The FLRW patch is then covered by the two static patches.  Let us note that in the two-dimensional case
a further third static patch (beyond the horizon) exists.

The  action for a conformally coupled scalar on de Sitter space reads  
\eq{
\label{action_conf_nd}
S_{m} = - \int d^{n}x \sqrt{- g}
\left[\frac{1}{2} \partial_{\mu} \Phi\, \partial^{\mu} \Phi + \frac{\xi}{2}R\,\Phi^2\right],
}
where $\xi = \frac{(n-2)}{4(n-1)}$ is the conformal coupling and
$R=n(n-1)H^2$ denotes the Ricci scalar of $dS_n$.
The object of interest is the vacuum expectation value $\langle T_{\mu\nu}\rangle$ of the
energy-momentum tensor
\eq{
  \label{EMtensor}
       T_{\mu\nu}=\partial_\mu \Phi \partial_\nu \Phi -{1\over
         2} g_{\mu\nu} \partial^\rho\Phi\partial_\rho\Phi
        +\xi \Big(G_{\mu\nu}-\nabla_\mu\nabla_\nu
          +g_{\mu\nu}\Box\Big) \Phi^2
}
in the Bunch-Davies vacuum. Here $G_{\mu\nu}$ denotes the
Einstein-tensor.  For clarity let us now focus on the
  two-dimensional case, where one can employ results from 2D conformal
  field theory\footnote{Of course 2D is special, as the Einstein
    tensor  is actually  vanishing  and pure gravity is
    topological. However, this point does not influence the following
    arguments about quantizing a scalar field on a classical curved 2D dS background.}.

\subsubsection*{Conformal scalar in 2D}

Since we are dealing with a 2D conformal field theory, it is
appropriate to introduce light-cone coordinates\footnote{The
  discussion of this simple case is largely influenced by the recent
  work \cite{Aalsma:2021kle}, in which the authors pointed out a
  misconception of \cite{Markkanen:2017abw}, on which the first version of our paper
  was based. We also thank the authors and in particular Lars Aalsma,
  Gary Shiu and Jan Pieter van der Schaar for clarifying discussions.}.
In the FLRW patch these are given by
\eq{
              U=\eta-x\,,\quad V=\eta+x \ \Rightarrow\
              ds^2=-{\textstyle {4\over H^2
                 (U+V)^2}} \, dU\, dV \,,
}
whereas in static coordinates one can introduce
\eq{
            u=\tau-r^*\,,\quad v=\tau+r^* \ \Rightarrow\
              ds^2=-{\textstyle {1\over
                 \cosh^2\left({H\over 2}(v-u)\right)}} \, du\, dv
}
with $r={1\over H} \tanh(Hr^*)$. These two metrics are conformally
equivalent via the corresponding conformal transformation inside the horizon
\eq{
        U=-{1\over H} e^{-H u}\,,\qquad V=-{1\over H} e^{-H v}\,,
}   
while similar expressions exist for the remaining patches\cite{Markkanen:2017abw}.  
In FLRW coordinates the vacuum expectation value of the
energy-momentum in the Bunch-Davies  vacuum can be computed as
\begin{equation}
  \langle T^{\rm div}_{UU}\rangle_{BD}=\langle T^{\rm div}_{VV}\rangle_{BD}={1\over 4\pi} \int_0^\infty k dk\,,\qquad
      \langle T^{\rm div}_{UV}\rangle_{BD}=0
\end{equation}
which is of course divergent and needs to be regularized.
Applying adiabatic regularization \cite{Parker:1974qw,Bunch:1980vc} gives the counterterms
\begin{equation}
\delta T^{BD}_{UU} =\delta T^{BD}_{VV} ={1\over 4\pi} \int_0^\infty k dk\,,\qquad \delta T^{BD}_{UV}={H^2\over 24 \pi} g_{UV}
\end{equation}
leading to the renormalized energy-momentum tensor
\begin{equation}
  \langle T^{\rm ren, FLRW}_{UU}\rangle_{BD}=\langle T^{\rm ren, FLRW}_{VV}\rangle_{BD}=0\,,\qquad
      \langle T^{\rm ren, FLRW}_{UV}\rangle_{BD}=-{H^2\over 24 \pi } g_{UV} \,.
 \end{equation}
 Note that the off-diagonal term is dictated by the general form of
 the 2D conformal anomaly  $\langle T^\mu_\mu\rangle =-{c\over 24\pi} R$.
Moreover, since $\langle T_{\mu\nu}\rangle \sim g_{\mu\nu}$ the
energy-momentum tensor preserves  the dS isometries and is
covariantly conserved, i.e. $\nabla^\mu \langle T_{\mu\nu}\rangle=0$.  

Next, we want to compute the quantum energy-momentum tensor in the static
patch. Since the static coordinates are related to the FLRW ones via
a conformal transformation, we can employ the anomalous transformation
law of the energy-momentum tensor,  which in general reads
\begin{equation}
                                \left({\partial U\over \partial
                                    u}\right)^2   T_{UU}(U)=T_{uu}(u) + {c\over
                                  24 \pi} \{U,u\}
\end{equation}
with $\{U,u\}$ denoting the Schwarzian
derivative and an analogous relation for the remaining light-cone
coordinates $V$ and $v$.
Here the ``unusual" plus sign in the right-hand side is due to the Lorentzian signature.
In this way we can derive
\begin{equation}
  \label{EMlocal2D}
  \langle T^{\rm ren, stat}_{uu}\rangle_{BD}=\langle T^{\rm ren,
    stat}_{vv}\rangle_{BD}={H^2\over 48\pi}\,,\qquad
      \langle T^{\rm ren, stat}_{uv}\rangle_{BD}=-{H^2\over 24 \pi } g_{uv} \,.
    \end{equation}
which still is covariantly conserved.
 Note that due to the non-vanishing conformal anomaly it  is
 inevitable that the final result is not proportional to the
 metric.  As we will see later, this is a general  behavior of the VEV
 of the energy-momentum tensor in the static patch.
 Therefore, this energy-momentum tensor is not behaving like
 a cosmological constant but has a component that contributes
 like radiation with the 2D equation of state $p=\rho$. 

As will further be discussed in the next section,  one can derive this result also directly 
by quantizing the scalar field in the static patch.  
After deriving the properly normalized solutions in regions A and B
(inside/outside of the horizon), it is possible to construct the two
continuous linear combinations 
$\Phi_A +\gamma (\Phi_B)^{*}$ and  $\gamma (\Phi_A)^{*} + \Phi_B$, 
where $\gamma=\exp(2\pi k/H)$ has the
form of  a thermal factor for the Gibbons-Hawking temperature $T=H/(2\pi)$.
From these combinations, the Bunch-Davies vacuum state can be
expressed as the entangled  so-called thermo-field double state over the two regions \cite{Goheer:2002vf}.
For a local  observable inside the horizon, one can trace over region
B resulting in a reduced (thermal) density matrix  $\hat \rho$ so that the
VEV of the energy-momentum tensor in region A can be computed  via
$\langle {T_A} \rangle = {\rm tr}(\hat{\rho} \,{T_A})$. 

In the conformal 2D case, one finds the divergent result
\eq{
 \label{staticdivT}
               &\langle T^{\rm div}_{uu}\rangle_{BD}=\langle T^{\rm
                 div}_{vv}\rangle_{BD}={1\over 4\pi} \int_0^\infty dk\,
               k 
                   \Big(1+{2\over e^{2\pi k/H}-1}\Big)\,,\\
               &\langle T^{\rm div}_{uv}\rangle_{BD}=0 .
             }
Next the issue of regularization arises. Here we cannot employ adiabatic regularization which is
taylor-made for fields slowly varying with  time. It goes beyond the scope of this paper to
develop a consistent theory of regularization in the static patch, but
we can make an important observation that will be sufficient for our purposes.
   
The divergent piece in \eqref{staticdivT} comes from the zero-point energy and can be cancelled
by using normal ordering among the modes. By doing this, one finds
that the remaining flat space   integral over the
Bose-Einstein distribution indeed gives $H^2/(48\pi)$, consistent
with \eqref{EMlocal2D}.  As is obvious from the conformal anomaly, there must also be an
off-diagonal counterterm $ \delta T^{BD}_{uv}$ which however is
proportional to the metric. Therefore, in this paper we work under the
following well-motivated mild assumption.
\begin{quotation}
  \noindent
  {\it Assumption 1: A non-trivial matter contribution  $\langle T^{\rm ren}_{\mu\nu}\rangle_{\rm M}$
   in the static patch can be detected by computing the initially  divergent vacuum
   expectation value of the energy-momentum tensor using the thermal
   reduced
   density matrix and then regularize it by normal ordering. }
\end{quotation}

\noindent
This in particular means that the other counterterms do not change
the fully renormalized energy-momentum tensor such that eventually
it preserves the dS isometries.
Note that the so obtained energy
density is time-independent so that the observer in
the static patch would attribute a temperature $T=H/(2\pi)$ to the horizon that
leads to a permanent inflow of Hawking radiation that is compensated by the
dilution and redshifting of the inflating dS space.

Transforming this result to the local coordinates $(\tau,r)$ one
can write
\eq{
  \langle T^\mu{}_\nu\rangle&= \langle T^\mu{}_\nu\rangle_{\rm M}+
   \langle T^\mu{}_\nu\rangle_{\rm CC} \\
  &={H^2\over 24\pi (1-H^2 r^2)}  {\rm diag}(-1,1)-{H^2\over
    24\pi} g^\mu{}_\nu
}
where we have indicated the two contributions with  equation
of state parameters $\omega_M=1$ and $\omega_{CC}=-1$.
This  form is often encountered in the early literature and  reveals
the appearance of a singularity at the horizon.

Similar computations have been performed for conformal scalar fields propagating
on higher dimensional de Sitter spaces. The behavior found in 2D
persists.  For instance in 4D we recall the textbook result from \cite{Birrell:1982ix}
\eq{
  \label{EM4Dconformal}
  \langle T^\mu{}_\nu\rangle={H^4\over 480\pi^2 (1-H^2 r^2)^2}  {\rm diag}\big(-1,{\textstyle{1\over
    3},{1\over 3},{1\over 3}}\big)+{H^4\over
    960\pi^2} g^\mu{}_\nu
}
where in the first term one also gets a flat space thermal integral
\eq{
     \int_0^\infty {d^3k\over (2\pi)^3} {k\over e^{2\pi k/H}-1}=
  \int_0^\infty {dk\over 2\pi^2} {k^3\over e^{2\pi k/H}-1}={H^4\over 480\pi^2}\,.
}
This energy-momentum tensor  is still time-independent, covariantly conserved and contains
%Moreover, there continues to be a contribution  to the static energy-momentum tensor
a contribution with the  equation of state of radiation $p=\rho/3$.

\subsection*{Remarks}

Let us finish this section with two  remarks.

\begin{itemize}
  \item{
The obtained energy-momentum tensors  are closely related to similar
results for Rindler space (see e.g. \cite{Davies:unknown,Bunch:1978ka,Hawking:1979ig,Candelas:1978gf}). 
As shown in \cite{Davies:unknown,Hawking:1979ig},  the FLRW patch of dS
is in the same conformal equivalence class as flat Minkowski space,
whereas the static patch of dS is in the same class as the Rindler
wedge. In particular, a conformal scalar on the Rindler wedge  also
gives rise to a contribution to $\langle T_{\mu\nu}\rangle$ with an
equation of state of radiation.}
\item{Using the results from the next section, we can also compute
    the energy-momentum tensor for a scalar in
    three-dimensional dS space. At the center of the static patch we
    obtain
    before regularization
\eq{
     \langle T^\mu{}_\nu \rangle_{r=0}=\int {d^2 k\over (2\pi)^2}
      k\tanh({\textstyle {\pi k\over H}})
                       \left({1\over 2}+{1\over e^{2\pi k/H}-1}\right)
      {\rm diag}\big(-1,{\textstyle{1\over
    2},{1\over 2}}\big)
}
 where the extra factor $\tanh(\pi k/H)$ arises from the
normalization \eqref{normalization_constant}
of the 3D wave function. It has the effect of introducing an extra
suppression of the  IR modes $k\ll H$. 
If we now regularize the integral by
subtracting the zero-point contribution $1/2$ we can write
\eq{
     \langle T^\mu{}_\nu\rangle_{r=0}&=\int {d^2 k\over (2\pi)^2} 
                       {k\over e^{2\pi k/H}+1}
      {\rm diag}\big(-1,{\textstyle{1\over
          2},{1\over 2}}\big)\\
        &={3 \zeta(3)\over 32 \pi^4} H^3{\rm diag}\big(-1,{\textstyle{1\over
    2},{1\over 2}}\big)
}
which intriguingly  looks like the thermal  expression for a free gas
of fermions. This behavior persists in all odd dimensions.
We cannot offer an  intuitive understanding of this
boson-fermion flip but note that for the quantum theory on dS
differences  between  even and odd dimensions have already
been observed e.g. in \cite{Bousso:2001mw,Lagogiannis:2011st}.
Note that the integrated result still shows the expected
$H^3$ scaling. 
With respect to our first remark, we have also checked that
the generalization of the 4D result for Rindler space from \cite{Candelas:1977zza}  to
3D leads to the same additional $\tanh(\pi k/H)$ factor.
}
\end{itemize}

\subsection{Proposal: Quantum Breaking}
\label{sec_quantumbreak}
The question is what do the results reviewed in the previous
subsection tell us about quantum gravity.  Clearly, this was just a
semiclassical analysis where a scalar field was quantized
on a classical curved manifold. The metric itself was not a quantum
object. The natural next step would be to include the computed
quantum $\langle T_{\mu\nu}\rangle$ on the right-hand side
of the Einstein equation and consider its backreaction.

As we have seen, while $\langle T^{\rm ren}_{\mu\nu}\rangle$ is still
covariantly conserved,  it is not transforming as a tensor under
general diffeomorphisms. The latter was evident  in the 2D conformal
case, where it could be related to the well known 2D conformal
anomaly. Thus, something is at  odds here, as roughly speaking general diffeomorphisms
play different roles for the left and the right-hand side of the Einstein
equation:
\begin{itemize}
\item{In general relativity general covariance is the {\it local}  symmetry
    principle and is not expected to be broken by quantum effects.}
\item{In the semi-classical conformal field theory, some of these
    diffeomorphisms are part of the {\it global} conformal symmetry, that
    can receive an anomaly.}
\end{itemize}

Thus, whether including the backreaction is a reasonable thing to do is a matter of debate,
as was expressed in the following statement from
the standard textbook 'Quantum Fields in Curved Space' by N.D. Birrell and
P.C.W. Davies, Cambridge Univ. Press 1982 \cite{Birrell:1982ix}:

\begin{quotation}
{\it  'When,   if   ever,   will   the   'back-reaction'
  (i.e. gravitational  dynamics  modified  by  gravitationally
  induced $\langle T_{\mu\nu}\rangle$)   be  approximately
  determined  by $T^{\rm ren}_{\mu\nu}$  computed  at  the  one-loop
  level?
  Misgivings about  these issues have been expressed by a number of
  authors.
  Many  of  them  might  be  resolved  if a full  theory  of quantum
  gravity  were  available,  to  which  one  could  claim that  the
  semiclassical  theory  is some  sort  of  approximation.'
}
\end{quotation}

Our point of view is that, since string theory is a theory of quantum gravity,
 its latest developments,  in particular the swampland program,  might shed some new light on this 
 long-standing problem.
 Let us first look at the BD vacuum in FLRW coordinates.  We have
seen that in $n$ dimensions $\langle T^{\rm ren}_{\mu\nu}\rangle\sim
g_{\mu\nu} H^n$ so it contributes
on the right-hand side of the Einstein equation like the cosmological
constant
\eq{
                        R_{\mu\nu}-{1\over 2} g_{\mu\nu}
                        R=\Lambda^{\rm cl}
                          g_{\mu\nu} +8\pi G\,\langle T^{\rm
                            ren}_{\mu\nu}\rangle_{\rm CC}\,
                        }
(in the remainder of this section we set the reduced Planck scale to one,
i.e. $(8\pi G)^{-1} =M_{\rm pl}^{n-2}=1$.                        
In FLRW coordinates this leads to the single  slightly perturbed Friedmann equation
\eq{
                    {{(n-1)(n-2)\over 2}} H^2 &=\Lambda^{\rm cl}+ \kappa H^n
}
which for $n>2$ and sub-Planckian energy scales just induces a small
shift of $H$ from its initial value. Therefore,  an observer in the
FLRW patch would not see any dramatic change,  as dS space would still
be a consistent solution to the backreacted Einstein
equation\footnote{Considering also perturbations around the background,
  there have been indications, see for instance
  \cite{Antoniadis:2006wq,Anderson:2013zia,Matsui:2018iez,
    Matsui:2019tlf,Chesler:2020exl}, that the backreaction leads to a
  more subtle picture where dS may be an unstable solution.}.

However, for an observer in the center of their static patch the
backreaction would be much more substantial due to the radiation-like component
in $T^{\rm ren}_{\mu\nu}$
\eq{
                        R_{\mu\nu}-{1\over 2} g_{\mu\nu} R=\Lambda^{\rm cl}
                          g_{\mu\nu} +\langle T^{\rm
                            ren}_{\mu\nu}\rangle_{\rm CC}+\langle T^{\rm
                            ren}_{\mu\nu}\rangle_{\rm M}\,.
}
In this case,  static de Sitter would no
longer be a consistent solution of the backreacted Einstein equation.
The additional component would cause $H$ to change with time, leading at best to a quasi dS space. 
However, this seems to be quite a  paradoxical situation:
observers in different patches would predict qualitatively 
different time evolution of space-time.

One way to resolve this is to say that the quantum energy-momentum
tensor $\langle T^{\rm ren}_{\mu\nu}\rangle$ in the static patch is
physically not reasonable,  as it for instance features a singularity at
the dS horizon and should therefore better not be included on the right-hand
side of the Einstein equation.

However, in view of the recent swampland conjectures a different
possibility is conceivable.  In order to resolve the paradoxical
situation, there should better not exist a static dS solution in quantum
gravity  in the first place. Thus, the best one can hope for is a quasi
dS solution, with for instance a rolling quintessence field involved.
In other words, there should necessarily always exist a non-vanishing classical
contribution $T^{\rm cl}_{\mu\nu}(\Phi)$ of the form \eqref{EMtensor},
induced by a gradient of
$\Phi$, on the right-hand side of the Einstein equation.
Hence,  we can group the  contributions on the right-hand side as follows
\eq{\label{Einstein_sym}
                        R_{\mu\nu}-{1\over 2} g_{\mu\nu} R=\Big(\Lambda
                          g_{\mu\nu} +\langle T^{\rm
                            ren}_{\mu\nu}\rangle_{\rm CC}\Big)+\Big(
                        T^{\rm cl}_{\mu\nu}(\Phi)+\langle T^{\rm
                            ren}_{\mu\nu}\rangle_{\rm M}\Big)\,.
                        }
%In  the FLRW patch the Friedmann equations would read 
%\eq{
%\label{friedmaneq}
%   {{(n-1)(n-2)\over 2}} H^2 &=\Lambda+{\dot\phi^2\over 2} \, ,\\[0.1cm]
%    -{{(n-1)(n-2)\over 2}} H^2 
%   -(n-2)\dot H &=-\Lambda+{\dot\phi^2\over 2} \, ,
 %}
% where $\dot\phi= 0$ is forbidden by the dS swampland conjecture.
% Note that  we have set $M_{\rm pl}=1$.
In the static patch, also the matter component  $\langle T^{\rm
  ren}_{\mu\nu}\rangle_{\rm M}$  could be just a small
correction to the classical rolling field energy.
Of course now the whole computation needs to be redone for the
  full  dynamical system, which is not easily feasible.
As a first rough but still quantitative estimate, 
requiring the naive dS quantum correction to be smaller than the
contribution from the 
classical,  time-dependent rolling field $\Phi(\tau)$ leads
to $\dot\Phi^2\gg \kappa H^n$.
Recalling  the break-time for a slowly rolling field as $t_Q\sim
1/(H\epsilon)$
and using the general relation for the slow-roll parameter
$\epsilon\sim (\dot\Phi)^2/H^2\sim H^{n-2}$ gives
\eq{
\label{quantumbreaktime}
            t_Q\sim {1\over H\epsilon} \sim {1\over H^{n-1}} \sim {M_{\rm pl}^{n-2}\over H^{n-1}
                      }
}                      
where in the last term we have reinstated the Planck scale.
Remarkably, this reasoning leads precisely to the proposed dS quantum
break time of Dvali-Gomez-Zell
\cite{Dvali:2013eja,Dvali:2014gua,Dvali:2017eba}.

Therefore, in the course of this  paper we follow the proposal:

\begin{quotation}
  \noindent
  {\it Assumption 2: The matter contribution  $\langle T^{\rm ren}_{\mu\nu}\rangle_{\rm M}$
is physically reasonable and is a mani\-festation  of quantum breaking
of dS space. Hence, it can be employed  as a method for  computing the
quantum break time. }
\end{quotation}

\noindent
Note that the scaling of
the quantum break time $t_Q$
with $H$ only depends on the power of $H$ in
$\langle T^{\rm
  ren}_{\mu\nu}\rangle_{\rm M}$, whereas
$\kappa$ only influences the numerical prefactor.
Identifying the cosmological constant with the value of a potential
$V,$ the bound for $\epsilon$ translates into
\eq{
      {|V'|\over V} \gtrsim c \,V^{n-2\over 4}
}
in natural units. As anticipated in the introduction, 
this is of the same form as  the no eternal inflation
bound of \cite{Rudelius:2019cfh}.

Let us also consider a dS maximum with a tachyonic instability, 
i.e. a quadratic hill-top potential.
Solving the Fokker-Planck equation for the quantum fluctuations of the
field around the maximum, leads to a life-time 
{$t_{\rm tac}\sim (n-1) H/|m^2|$}.
Requiring that the quantum break time is longer than
this lifetime, leads to the bound  ${|V''|/V} \gtrsim c'\, V^{n-2\over 2}$,
which is actually  weaker than the no eternal inflation bound
\cite{Rudelius:2019cfh}.  However,  as already noted in \cite{Dvali:2017eba}, one should keep in mind that the system only decays if it is not
stuck in eternal inflation. This means that $t_{\rm tac}$ only
is the correct life-time if the tachyonic decay is not beaten by the competing exponential
expansion of space, i.e. we also need to require $t_{\rm tac}\lesssim 1/H$.  Therefore, we  cannot get
a bound from quantum breaking that is weaker than the one for
no eternal inflation. As a consequence, the bound for a dS maximum
is given by  
\eq{
  {|V''|\over V} \gtrsim c'\,.
}
This  scales in the same way as the second clause of the refined swampland conjecture, provided 
$c'$ is of appropriate value.
For the TCC, a $\log$-corrected\footnote{The appearance of $\log$-corrections for swampland conjectures was also studied in \cite{Blumenhagen:2019vgj,Blumenhagen:2020dea}.} bound of the same type was
derived \cite{Bedroya:2019snp}, that essentially scales as
\eq{
\label{TCCtachy}
                  {|V''|\over V} > {16\over (n-1)(n-2)}\, \left(\log
                   V\right)^{-2}\,.
}
Throughout this paper we are ignoring such $\log$-correcti\-ons. 

\subsubsection*{Remarks}

Let us conclude this section with two remarks:

\begin{itemize}
 \item{Strictly following this logic we are led to an immediate conclusion
that at first sight seems fairly disturbing but  at second sight might
be not.
As we have commented on,  a scalar field on a Rindler wedge also features
a non-vanishing quantum energy-momentum tensor so that we would
conclude that  eternal Minkowski space cannot be a consistent
solution of quantum gravity, either. The question of the existence
of Minkowski minima in string theories with at most $N=1$
supersymmetry has been addressed recently by Palti,Vafa, Weigand
\cite{Palti:2020qlc}, where it was argued that such vacua are not in the string
landscape. Clearly, this is an important question that needs to be
addressed further in the future. However, here we now continue
under the working assumption that matter-like
contributions in $\langle T^{\rm ren}_{\mu\nu}\rangle$ indicate
quantum breaking and discuss further consequences.} 
\item{The question  we  eventually want to address  is whether 
censoring the quantum breaking process can also directly
lead to the bound on $|V'|/V$  from the (refined) dS swampland conjecture.
Apparently, this would require a change in the scaling of
$\langle T^{\rm  ren}_{\mu\nu}\rangle_{\rm  M}$ with
respect to  the Gibbons-Hawking temperature. 
For that purpose we have to include more effects from
quantum gravity into the game, i.e. we have to go beyond
the semiclassical analysis and ideally perform the analysis
in the full string theory. This is 
not straightforward, but the simple flat space thermal 
contribution we found calls for a natural generalization to string
theory.
Thus, we are led to consider strings at finite temperature
${T=H/(2\pi)}$.}
\end{itemize}

\subsection{Reduced density matrix for massive scalar}
\label{sec_Markkanen_generalization}

Due to the simple thermal interpretation of the result in the conformal case, it is
reasonable that it continues to  hold  for the massive
case as well. This is what we shall examine next. 
If this is indeed the case, we will have a solid starting point for a
generalization to string theory, since in Einstein frame all (massive)
string excitations are minimally coupled to gravity, i.e. $\xi=0$ in
\eqref{action_conf_nd}. 

\subsubsection*{Solutions to the equations of motion}

Here we closely follow the calculation presented
in\cite{Markkanen:2017abw}
for the massive scalar in an $n$-dimensional 
dS space-time, described by 
\eq{
\label{action_m_nd}
S_{m} = - \int d^{n}x \sqrt{- g}
\left[\frac{1}{2} \partial_{\mu} \Phi\,\partial^{\mu} \Phi +
  \frac{\xi}{2} R \Phi^2 +\frac{m^2}{2}\Phi^2\right],
}
where we also allowed a general coupling to the dS curvature
$R=n(n-1) H^2$. Recall that for $m=0$ and $\xi={(n-2)\over 4(n-1)}$ we have the
conformal setting.  Scalar field quantization in static dS space has also been discussed in e.g.
\cite{Polarski:1990tr,Polarski:1990ux}.
 
We start by solving the corresponding equation of motion 
\eq{
\label{eomscalar}
       \Big(\Box - m^2-\xi n(n-1) H^2\Big)\Phi = 0\,.    
     }
In FLRW coordinates the solution can be expanded as
\eq{
          \Phi=\int {d^{n-1} {\bf k}\over \sqrt{(2\pi a)^{n-1}}}\,
          e^{i{\bf kx}}
          \left( \hat a_{\bf k}\, f_{\bf k}(t) + \hat a^\dagger_{-{\bf
                k}}\, f^*_{\bf k}(t)\right) ,
}
where the modes satisfy
\eq{
              \ddot f_{\bf k}(t) +\omega^2_{\bf k}\, f_{\bf k}(t)=0
}
with
\eq{    \omega_{\bf k}^2={{\bf k}^2\over a^2} + H^2 \gamma^2\,,\qquad
      \gamma=\left( {m^2\over H^2} +\xi n(n-1) -{(n-1)^2\over 4} \right)^{1\over 2}\,.
}
Taking the $t\to -\infty$ limit of the solution, given in terms of the
Hankel function as
\eq{
               f_{\bf k}(t)=\sqrt{\pi\over 4H} e^{-\pi\gamma/2}
               H^{(1)}_{i\gamma}\left( {|{\bf k}|\over aH}\right)\,,
}
reveals that it satisfies  Bunch-Davies boundary conditions.
Thus, the Bunch-Davies vacuum is defined via 
\eq{   
                      a_{\bf k} |0\rangle_{BD}=0\,.
}
This solution holds inside and outside of the dS horizon.
The task now is to find the corresponding solutions in the static
patches and finally express the Bunch-Davies vacuum as
an entangled state of states inside and outside the horizon.

For that purpose, one first has to solve the equation of motion
in static coordinates \eqref{static_de_Sitter}.
Variation of the action leads to the differential equation in the static coordinates 
\eq{
\label{KG_static_de_Sitter}
&−\frac{1}{r^{n-2}}\del_{\tau}\left( r^{n-2}\Big(\frac{1}{1−H^2r^2} \Big) \del_{\tau} \Phi \right)+\frac{1}{r^{n-2}}\del_{r}  \Big(r^{n-2}(1−H^2r^2) \,\del_{r} \Phi \Big)+ \\[0.1cm]
&\phantom{aaaaaaaaaaaaaaaaaaaaaaaa}+ \frac{1}{r^2} \nabla^{2}_{n-2} \Phi -\Big(m^2+\xi n(n-1)H^2\Big) \Phi = 0\,,    
}
which can be separated into radial, temporal and angular components with the ansatz
\eq{
\label{scalar_ans}
\Phi =N_{L \omega}\, f_{L \omega}(r)\,Y_{L, l_{1}, \dots, l_{n-3}}(\theta)\,e^{-i \omega \tau}\,.    
}
$Y_{L, l_{1}, \dots, l_{n-3}}(\theta)$ denote the (hyper-)spherical
harmonics in $(n-2)$-dimensions \cite{doi:10.1142/10690} and $f_{L
  \omega}(r)$ is  the solution to the radial differential equation
\eq{
\label{radialeom}
 & \frac{(1-H^2r^2)}{r^{n-2}} \del_r\Big(r^{n-2}(1-H^2r^2)\del_r  f_{L \omega}(r)\Big)+  \\[0.2cm]
&\bigg[\omega^2-(1-H^2r^2)\left(\frac{L(L+n-3)}{r^2}
  + \big(m^2+\xi n(n-1)H^2\big)\right) \bigg] f_{L \omega}(r) = 0\,.      
}
Here $-L(L+n-3)$ is the eigenvalue of the Laplace operator on
$S^{n-2}$ for $Y_{L, l_{1}, \dots, l_{n-3}}$. 
From here on we abbreviate the non-principal indices $ l_{1}$  to $ l_{n-3}$ as $ \lambda$.
Equation \eqref{radialeom} is simply a  hypergeometric differential equation with the  solution \cite{Abdalla:2002hg}
\eq{
\label{sol_radialeom}
 f_{L \omega}(r) = (Hr)^{L}\,&[1-(Hr)^2]^{\frac{i\omega}{2H}} \,\times\\[0.1cm]
 & {}_2 F_1 \left[ \textstyle\frac{1}{2}(L+ \textstyle\frac{i\omega}{H} +\mu_{-}), \textstyle\frac{1}{2}(L + \textstyle\frac{i\omega}{H} +\mu_{+});L+\textstyle\frac{n-1}{2};(Hr)^2 \right] ,     
 }
\noindent
where
\eq{
  \mu_{\pm} = \frac{1}{2}\left ((n-1) \pm \sqrt{(n-1)^2 -4\xi n(n-1)- 4m^2/H^2}
  \right)\,.
}  
The resulting mode expansion of the field operator $\Phi$  reads
\eq{
\label{scalar_exp}
\Phi &=\sum_{L,\lambda} \int_{0}^{\infty} d\omega \, \bigg[ N_{L
  \omega}\,(Hr)^{L}\,[1-(Hr)^2]^{\frac{i\omega}{2H}} \; Y_{L, \lambda}(\theta) \,e^{-i \omega \tau}\, \times \\
&\quad {}_2 F_1 \left[ \textstyle\frac{1}{2}(L+ \textstyle\frac{i\omega}{H}
  +\mu_{-}), \textstyle\frac{1}{2}(L + \textstyle\frac{i\omega}{H}
  +\mu_{+});L+\textstyle\frac{n-1}{2};(Hr)^2 \right] \,\hat{a}_{L
  \lambda \omega} +
{\rm H.C.} \bigg],
} 
where H.C. denotes the Hermitian conjugate. 
For the complete solution we still have to determine the normalization
constant $N_{L \omega}$, which we obtain by following the quantization
procedure of \cite{Higuchi:1986ww} and demanding that the commutation
relations  for $\Phi$
\eq{
\label{app_commrelPhi1}
[\Phi(r,\theta, \tau), &\dot{\Phi}(r',\theta', \tau)] = 
- \frac{i}{g^{\tau \tau}\sqrt{-g}} \,\delta(r-r')\,
\delta^{(n-2)}(\theta- \theta'), \\[0.1cm]
[\Phi(r,\theta, \tau), &\Phi(r',\theta', \tau)] =  [\dot{\Phi}(r,\theta, \tau), \dot{\Phi}(r',\theta', \tau)] = 0
}
and for the creation and annihilation operators 
\eq{
\label{app_commrela}
[\hat{a}_{L \lambda \omega}, \hat{a}^{\dagger}_{L' \lambda' \omega'}] &=   \delta(\omega-\omega') \,\delta_{LL'}\,\delta_{\lambda \lambda'}, \\
[\hat{a}_{L \lambda \omega}, \hat{a}_{L' \lambda' \omega'}] &= [\hat{a}^{\dagger}_{L \lambda \omega}, \hat{a}^{\dagger}_{L' \lambda' \omega'}] = 0 
}
are indeed fulfilled.
Following the steps laid out in appendix \ref{app_A}, the
normalization constant can be determined as
\eq{
\label{normalization_constant}
|N_{L \omega}|^2 = \frac{H^{n-2}}{4 \pi \omega }\,
\frac{\big|\Gamma\big(\frac{1}{2}(L + \frac{i\omega}{H}
  +\mu_{+})\big)\big|^2\; \big|\Gamma\big(\frac{1}{2}(L + \frac{i\omega}{H} +\mu_{-})\big)\big|^2} 
{\big|\Gamma\big(L + \frac{n-1}{2}\big)\big|^2\;
  \big|\Gamma\big(\frac{i\omega}{H}\big)\big|^2}\,.
}
After having derived the solution to the equation of motion of an
$n$-dimensional massive scalar field,  we can continue along the lines of \cite{Markkanen:2017abw}. 
Adopting the notation of the aforementioned paper, the  solution we just presented is the solution for the region A, which is the region inside the horizon. 
In the region B $-$ outside of the horizon $-$ the radial coordinate and
time coordinate exchange their roles as can 
be seen from the relation
\eq{
\label{radial_vector_regB}
\del_{r_B} = \frac{e^{-Ht}}{(Hr_B)^2 - 1} \big(H r_B\,e^{Ht} \, \del_{t} - \del_r\big),
}
where $t$ and $r$ denote the time and radial coordinates in the FLRW patch.
The inner product outside of the horizon then becomes
 \eq{
\label{inner_product_B}
(\Phi_1, \Phi_2)_B = -i \int d\Omega \int_{-\infty}^{\infty} d\tau_B \; r_B^{n-2}(H^2r_B^2 - 1) \,\Phi_1 \overleftrightarrow{\nabla}_{r_B} \Phi_{2}^{*}  \,.
} 
Since   it is independent of the choice of $r_B$,  this expression can
be evaluated at 
$H r_B \rightarrow 1$. 
With the inner product and the equation of motion of the scalar field, we can deduce the correct solution for region $B$
\eq{
\label{scalar_exp_regB}
\Phi^B_{L, l_{1} \dots l_{n-3}, \omega} = &N_{L \omega}^{*}\,
(Hr_B)^{L}\, [(Hr_B)^2 - 1]^{-\frac{ i\omega}{2H}}\, Y_{L,
  \lambda}^{*}(\theta)\, e^{i \omega \tau}\, \times \\[0.1cm]
& {}_2 F_1 \left[ \textstyle\frac{1}{2}(L- \textstyle\frac{i\omega}{H} +\mu_{-}), \textstyle\frac{1}{2}(L - \textstyle\frac{i\omega}{H} +\mu_{+});L+\textstyle\frac{n-1}{2};(Hr_B)^2 \right] \,.
}

\subsubsection*{The reduced density matrix}

In order to determine the reduced density matrix we first determine  the limits of
$\Phi^A_{L,\lambda, \omega}$ and $\Phi^B_{L, \lambda, \omega}$ at the
horizon and  construct a linear combination of $\Phi^A_{L, \lambda,
  \omega}$ and $\Phi^B_{L, \lambda, \omega}$, which is continuous
across the horizon.
For the limits we find
\eq{
\label{scalar_lim_regA/B}
\Phi^A_{L, \lambda, \omega} &\xrightarrow{Hr_{A} \rightarrow 1}
H^{n-2\over 2} \,\frac{Y_{L, \lambda}}{\sqrt{4 \pi \omega}} \big(\cosh^{-2}(H r_{A}^{*})\big)^{i \frac{\omega}{2H}} \,e^{-i\omega \tau_A}  \\
&\sim \frac{Y_{L, \lambda}}{\sqrt{4 \pi \omega}\, r_{A}^{n-2\over 2}} \,e^{-i \omega (\tau_{A} +  r_{A}^{*})}\,,\\
\\
\Phi^B_{L, \lambda, \omega} &\xrightarrow{Hr_{B} \rightarrow 1} H^{n-2\over 2}\, \frac{(Y_{L, \lambda})^{*}}{\sqrt{4 \pi \omega}} \big(\sinh^{-2}(H r_{B}^{*})\big)^{-i \frac{\omega}{2H}} \,e^{i\omega \tau_B}  \\
&\sim \frac{(Y_{L, \lambda})^{*}}{\sqrt{4 \pi \omega}\,r_{B}^{n-2\over 2}} \,e^{i \omega (\tau_{B} +  r_{B}^{*})}\,,
}
where $r_{A}^{*}$ denotes the tortoise coordinate defined as 
$r_{A}=H^{-1} \tanh(H r_{A}^{*})$ in region A. The tortoise
coordinate in region B is defined as $r_{B}=H^{-1} \coth(H r_{B}^{*})$.
This makes it evident that at the horizon the solutions become plane
waves with Bunch-Davies boundary conditions.
It turns out that the following linear combinations are continuous across the horizon
\eq{
\label{lin_combAB}
\Phi^A_{L, \lambda, \omega} + \gamma\, (\Phi^B_{L,\lambda, \omega})^{*} \quad \textrm{and} \quad  \gamma\, (\Phi^A_{L, \lambda, \omega})^{*} + \Phi^B_{L, \lambda, \omega}\,,
}
with $\gamma \equiv e^{- \frac{\pi \omega}{H}}$.  In order to see
this, one writes
\eq{
\label{lin_combAB2}
&\text{Region A:} \quad e^{-i \omega (\tau_{A} +  r_{A}^{*})} = e^{\frac{i \omega}{H} \log(- HV)}\,,\\
&\text{Region B:}\quad e^{- \frac{\pi \omega}{H}}\left(e^{i \omega (\tau_{B} +  r_{B}^{*})}\right)^* =e^{- \frac{\pi \omega}{H}} e^{\frac{i \omega}{H} \log(HV)}=e^{\frac{i \omega}{H} \log(- HV)}\,,
}
where  a branch cut such that  $\log(-1) = i\pi$ is used and the FLRW based
light-cone coordinate $V$ is  given in the static regions $A/B$ as
\eq{
\label{def_lc_V}
\text{Region A:}\  \;V = - \frac{1}{H} e^{-H(\tau_{A} +  r_{A}^{*})}\,,\qquad
\text{Region B:}\   \;V =  \frac{1}{H} e^{-H(\tau_{B} +  r_{B}^{*})}\,.
}

In fact the presented expressions \eqref{lin_combAB} are identical to
those in \cite{Markkanen:2017abw} except for the replacement of the
spherical harmonics with their generalization to higher
dimensions. As a consequence, in the $n$-dimensional massive case we
get analogously to \cite{Markkanen:2017abw} the following relations for the BD vacuum
\eq{
\label{vacuum_AB}
\Big(\hat{a}^A_{L \lambda \omega} - \gamma\, (\hat{a}^{B}_{L \lambda \omega})^{\dagger}\Big)\big|0_{L \lambda \omega}\big\rangle_{BD} = 0\,,\quad
\Big(\hat{a}^B_{L \lambda \omega} - \gamma\, (\hat{a}^{A}_{L \lambda \omega})^{\dagger}\Big)\big|0_{L \lambda \omega}\big\rangle_{BD} = 0\,.
}
The properly normalized solution to \eqref{vacuum_AB} is
\eq{
\label{sol_vacuum_AB}
|0_{L \lambda \omega}\rangle_{BD} = \sqrt{1-\gamma^{2}} \displaystyle
\sum_{n_{L\lambda \omega} = 0}^{\infty} \gamma^{n_{L\lambda
    \omega}} \, |n_{L\lambda \omega}, A\rangle\otimes |n_{L\lambda\omega}, B\rangle\,.
}
Thus, we have written the Bunch-Davies vacuum as a linear combination
of entangled states in the product Hilbert space ${\cal H}_A\otimes{\cal
  H}_B$. This is still a pure state but has the appropriate form to explicitly
carry out the trace over the unobservable states in region $B$.
In this way, we obtain the  reduced density matrix of the resulting mixed state
\eq{
\label{density_matrix}
\hat \rho = \displaystyle \prod_{L\lambda\omega} (1 -  e^{- \frac{2 \pi \omega}{H}}) \displaystyle \sum_{n_{L\lambda\omega} = 0}^{\infty} e^{- \frac{2 \pi \omega}{H} n_{L\lambda\omega}}\, |n_{L\lambda\omega}, A\rangle \langle n_{L\lambda\omega}, A|\,.
}
This has the form of a thermal state of temperature $T=H/(2\pi)$, which
is nothing else than the Gibbons-Hawking temperature of the dS horizon.

 \subsection{Energy-momentum tensor for massive scalar}
\label{sec_emtensor}

Now we are in a position to determine the   energy
momentum tensor,  in particular the energy density and the pressure
\eq{
\rho = \langle T_{00} \rangle = {\rm tr}(\hat \rho\, T_{00})\,,\qquad
   p = \langle T_{rr} \rangle = {\rm tr}(\hat \rho\, T_{rr})\,.
 }
Evaluating these one can reproduce the results for the conformally
coupled scalar with $m=0$ and $\xi={(n-2)\over 4(n-1)}$ presented in the
previous section. 
As an illuminating  non-conformal example, we consider the 4D massive conformally
coupled scalar field with $\xi=1/6$.
The energy-momentum tensor in this case can be expressed as
\eq{
  \label{Tnonconf4D}
  T_{\mu\nu}=\partial_\mu \Phi \partial_\nu \Phi-{1\over
    6}g_{\mu\nu}\,\partial^\rho \Phi\partial_\rho \Phi+{1\over 6} g_{\mu\nu} (H^2-m^2) \Phi^2
  -{1\over 6}\nabla_\mu\nabla_\nu (\Phi^2)
}
and we find $\mu_\pm={3\over 2}\pm {1\over 2}\sqrt{1-4{m^2\over
    H^2}}$.
Since we are interested in what observers at the center of their static
patch see, we work at leading order in $O(Hr)$.
At  this order, only the modes $L=0,1$ are relevant and have the form
\eq{
\label{zero_mode_approx}
\Phi_{0,\lambda = 0, \omega} = N_{0 \omega}\,  Y_{0, \lambda = 0}(\theta)\,
e^{- i \omega \tau} +  O(Hr)
}
and 
\eq{
\label{one_mode_approx}
\Phi_{1,\lambda, \omega} = N_{1 \omega}\, (Hr)\, Y_{1, \lambda}(\theta) \, e^{-
  i \omega \tau} + O(Hr)\,,\qquad \lambda\in\{-1,0,1\}\,.
}
Now we compute  the expectation value of the individual components of
$\rho=T_{00}$ and $p=T_{rr}$. The resulting expressions are listed in
appendix \ref{app_B}.  In contrast to
the conformal case, 
these cannot be evaluated analytically due to their dependence on a non-integer valued
mass.  For $T_{00}$ the last term in \eqref{Tnonconf4D} does not contribute
so that the regularized final result reads
\eq{
  \label{rhofinal4D}
  \rho=&\int_0^\infty d\omega \bigg[ {1\over 12\pi^4} (5 H\omega^2
  -H^3+ H m^2 )
                        \big|\Gamma\big({\textstyle {i\omega\over 2H}+{\mu_+\over
                          2}}\big)\big|^2
                         \big|\Gamma\big({\textstyle {i\omega\over
                             2H}+{\mu_-\over 2}}\big)\big|^2\\[0.1cm]
                          &+{1\over 3\pi^4} H^3
                        \big|\Gamma\big({\textstyle {i\omega\over 2H}+{\mu_+\over
                          2}+{1\over 2}}\big)\big|^2
                         \big|\Gamma\big({\textstyle {i\omega\over
                             2H}+{\mu_-\over 2}+{1\over
                             2}}\big)\big|^2 \bigg]\times
                          {\sinh({\pi \omega\over H})\over e^{2\pi\omega/H}-1} \,.
                        } 
 Following our assumption about regularization, we employed normal
 ordering    and removed the zero-point energy.  Recall that we expect
 more  counterterms to be present, which however do not make
 the final result proportional to the metric.  This assumption
 clearly holds for the conformal case and should also hold after
 turning on the continuous mass parameter $m$.
 
\begin{figure}[ht]
  \centering
  \includegraphics[width=8.5cm]{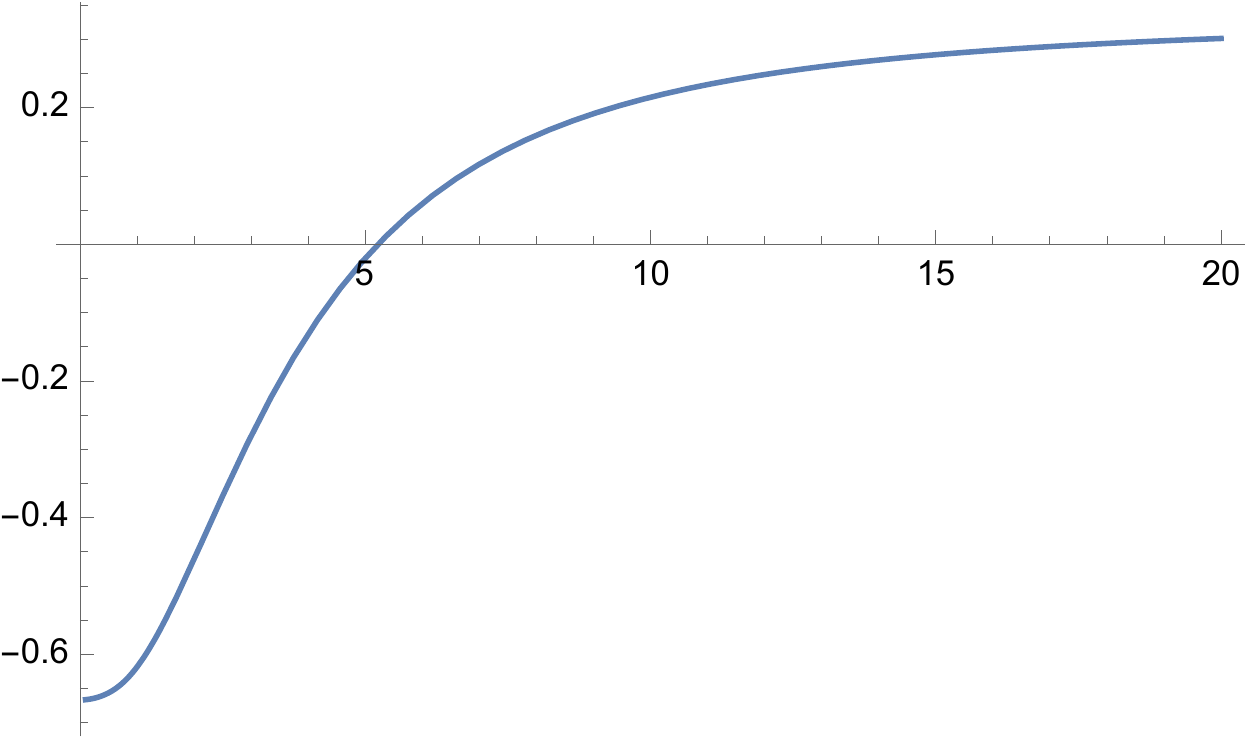}
\begin{picture}(0,0)
  \put(0,92){$H$}
   \put(-232,150){$\omega$}
   \end{picture}
  \caption{The dependence $\omega(H)$ at fixed value of the mass $m=2$.}
  \label{fig:wH}
\end{figure}
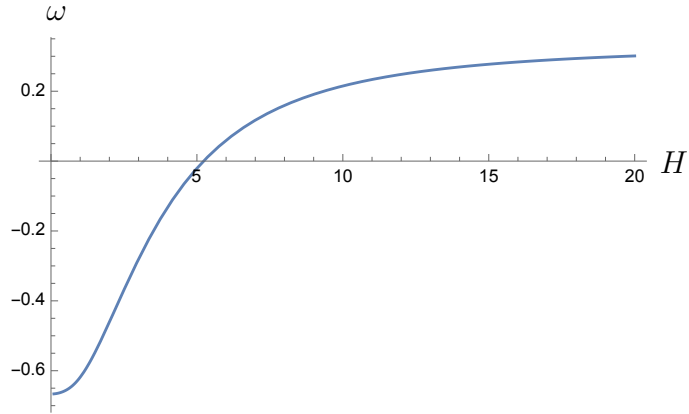

Indeed, for $m=0$ one gets $\mu_+=2$ and $\mu_-=1$ so that the contributions
from the $\Gamma$-functions can be evaluated analytically
\eq{
                 &\big|\Gamma\big({\textstyle {i\omega\over 2H}+{1\over
                          2}}\big)\big|^2
                         \big|\Gamma\big({\textstyle {i\omega\over
                             2H}+1}\big)\big|^2= {\pi^2 \omega\over H}
                         {1\over \sinh({\pi \omega\over H})}\\[0.1cm]
                  &\big|\Gamma\big({\textstyle {i\omega\over 2H}+{1}}\big)\big|^2
                         \big|\Gamma\big({\textstyle {i\omega\over
                             2H}+{3\over 2}}\big)\big|^2= {\pi^2 \omega\over H}
                        \Big({\omega^2\over 4H^2}+{1\over 4}\Big) {1\over \sinh({\pi \omega\over H})}
                      }
giving in total the energy-density at $r=0$ for the 4D conformal scalar from \eqref{EM4Dconformal}.
For the $T_{rr}$ component, the last term in \eqref{Tnonconf4D} can be written as
\eq{
  \nabla_r\nabla_r (\Phi^2)=2 (\partial_r\Phi)^2+2\Phi\partial_r^2
  \Phi\,\quad {\rm with}\quad 
  \partial_r^2 \Phi|_{r=0}={1\over 3}(2H^2+m^2-\omega^2)\,.
}  
This leads to the pressure
\eq{
   \label{pfinal4D}
  p=&\int_0^\infty d\omega \bigg[ {1\over 36\pi^4} (5 H\omega^2
  -H^3- 5 H m^2 )
                        \big|\Gamma\big({\textstyle {i\omega\over 2H}+{\mu_+\over
                          2}}\big)\big|^2
                         \big|\Gamma\big({\textstyle {i\omega\over
                             2H}+{\mu_-\over 2}}\big)\big|^2\\[0.1cm]
                          &+{1\over 9\pi^4} H^3
                        \big|\Gamma\big({\textstyle {i\omega\over 2H}+{\mu_+\over
                          2}+{1\over 2}}\big)\big|^2
                         \big|\Gamma\big({\textstyle {i\omega\over
                             2H}+{\mu_-\over 2}+{1\over
                             2}}\big)\big|^2 \bigg]\times
                          {\sinh({\pi \omega\over H})\over e^{2\pi\omega/H}-1} 
                       }   
which for $m=0$ agrees with the conformal result shown in
\eqref{EM4Dconformal}, as well.
However, in the massive case these integrals can only be evaluated
numerically. In figure \ref{fig:wH}, at fixed value of the mass $m$
we show the equation of state parameter
$\omega=p/\rho$ as a function of the Hubble constant $H$ (or
alternatively the Gibbons-Hawking temperature).

\vspace{0.3cm}
%%%%%%%%%%%%
%%%%%%%%%%%%
%%%%%%%%%%%%

Apparently, in the high temperature regime $H\gg m$ one finds
the conformal value $\omega=1/3$, whereas in the opposite
low temperature limit  $H\ll m$ one gets $\omega=-2/3$. Note that
this deviates from the result for a free  gas in flat space, which
has $\omega=0$ in the latter  limit.
In appendix \ref{app_C} we will see the origin of  the value $\omega=-2/3$  in the
$m/H\to \infty$ limit.

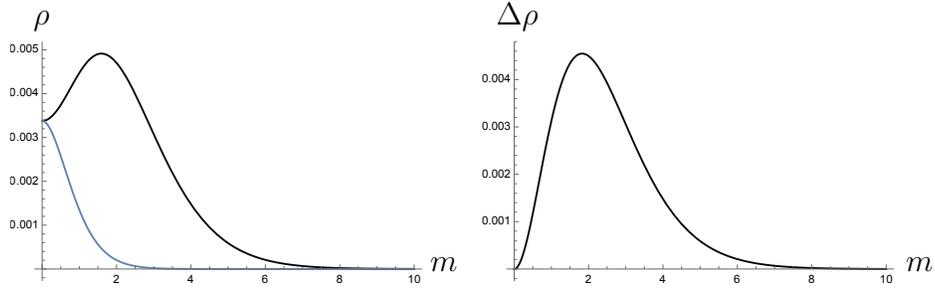
\begin{figure}[ht]
  \centering
  \includegraphics[width=5.5cm]{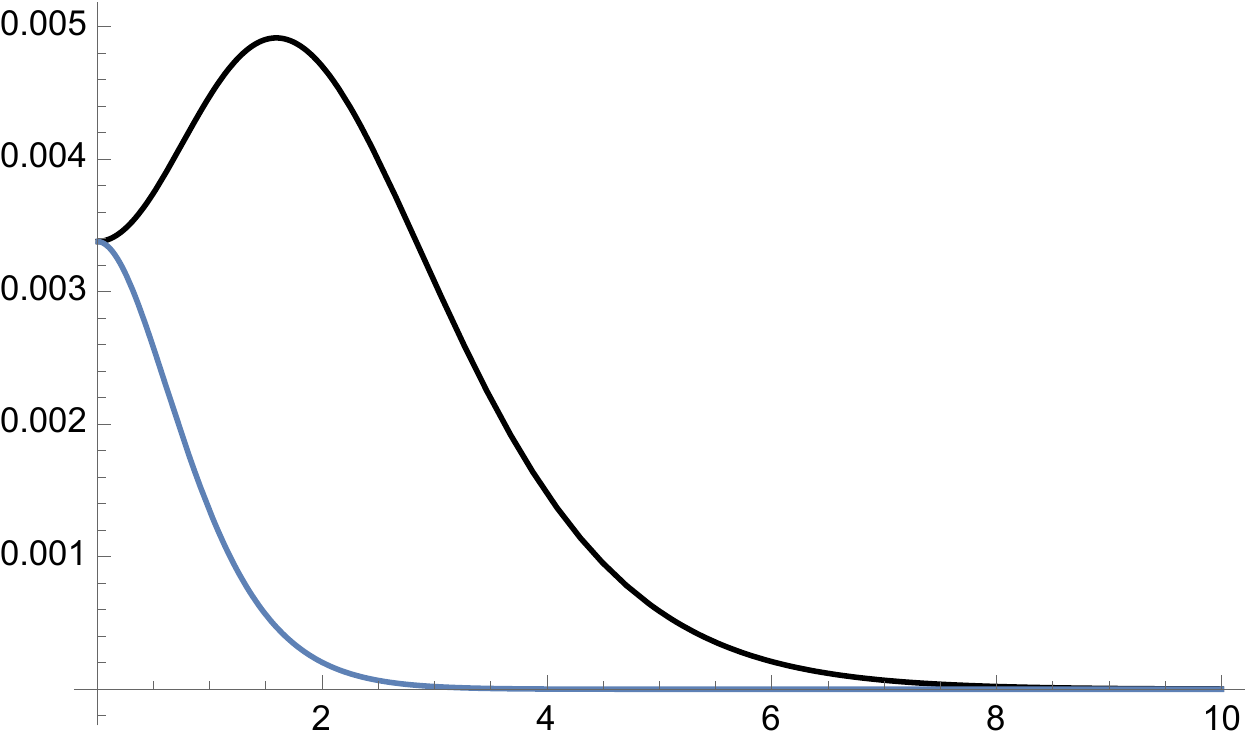}
   \hspace{0.5cm}
  \includegraphics[width=5.5cm]{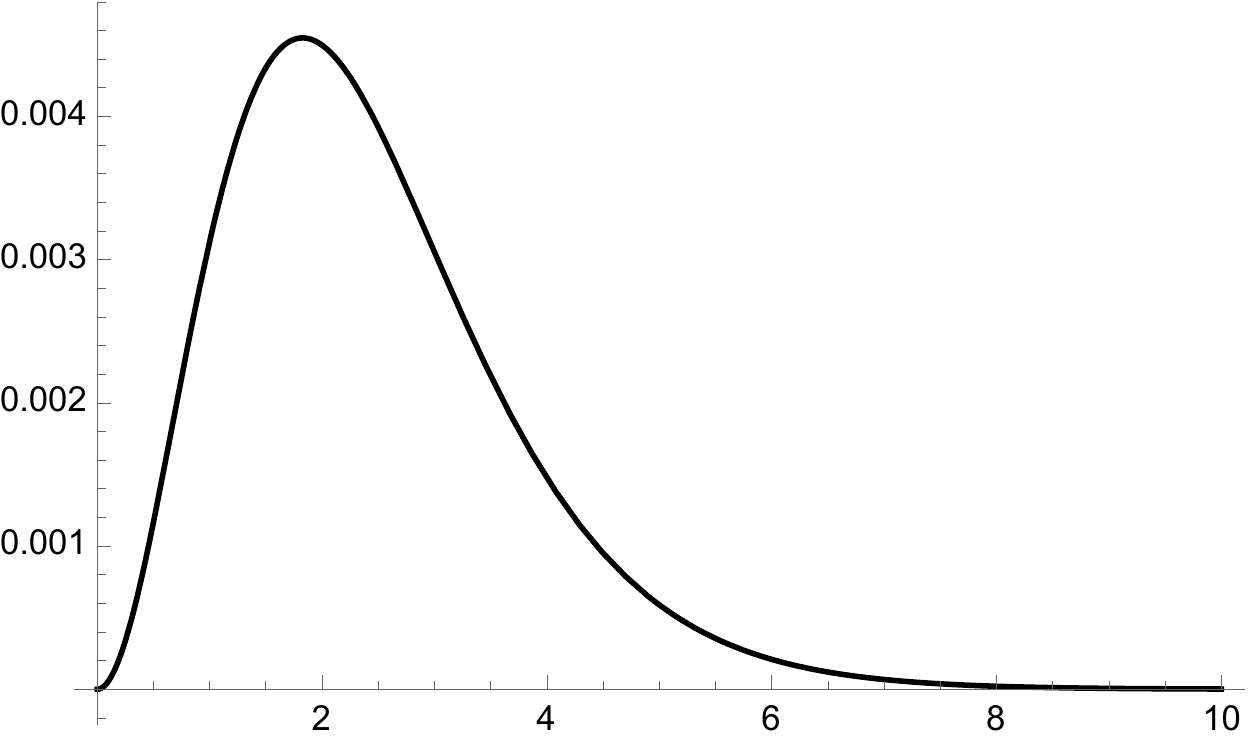}
\begin{picture}(0,0)
  \put(0,7){$m$}
   \put(-180,7){$m$}
   \put(-330,100){$\rho$}
   \put(-155,100){$\Delta\rho$}
   \end{picture}
  \caption{One the left: The dependence $\rho(m)$ at fixed value of the temperature
    $H=2$. The black curve is the expression \eqref{rhofinal4D},
    whereas the blue curve the flat space result
    \eqref{rhofreegas}. On the right: The difference of the two curves.}
  \label{fig:rhom}
\end{figure}

\begin{figure}[ht]
  \centering
  \includegraphics[width=5.5cm]{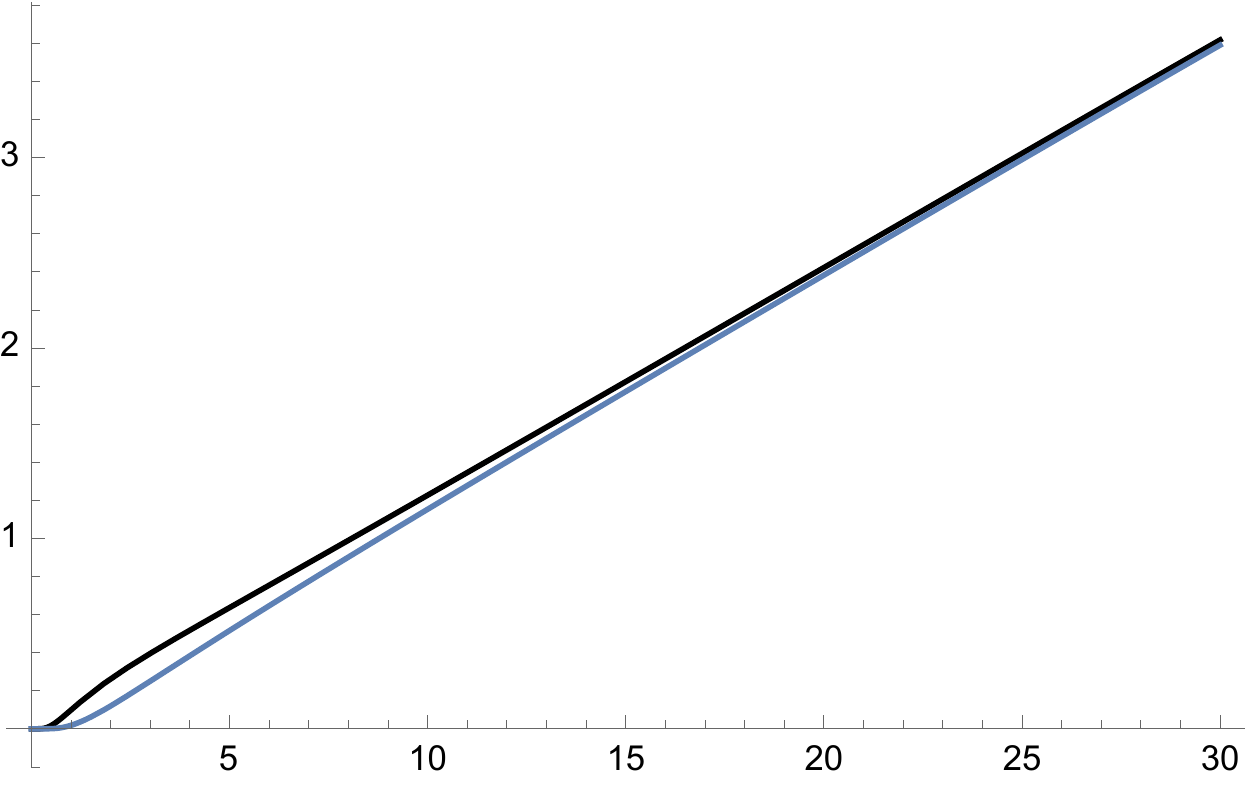}
   \hspace{0.5cm}
  \includegraphics[width=5.5cm]{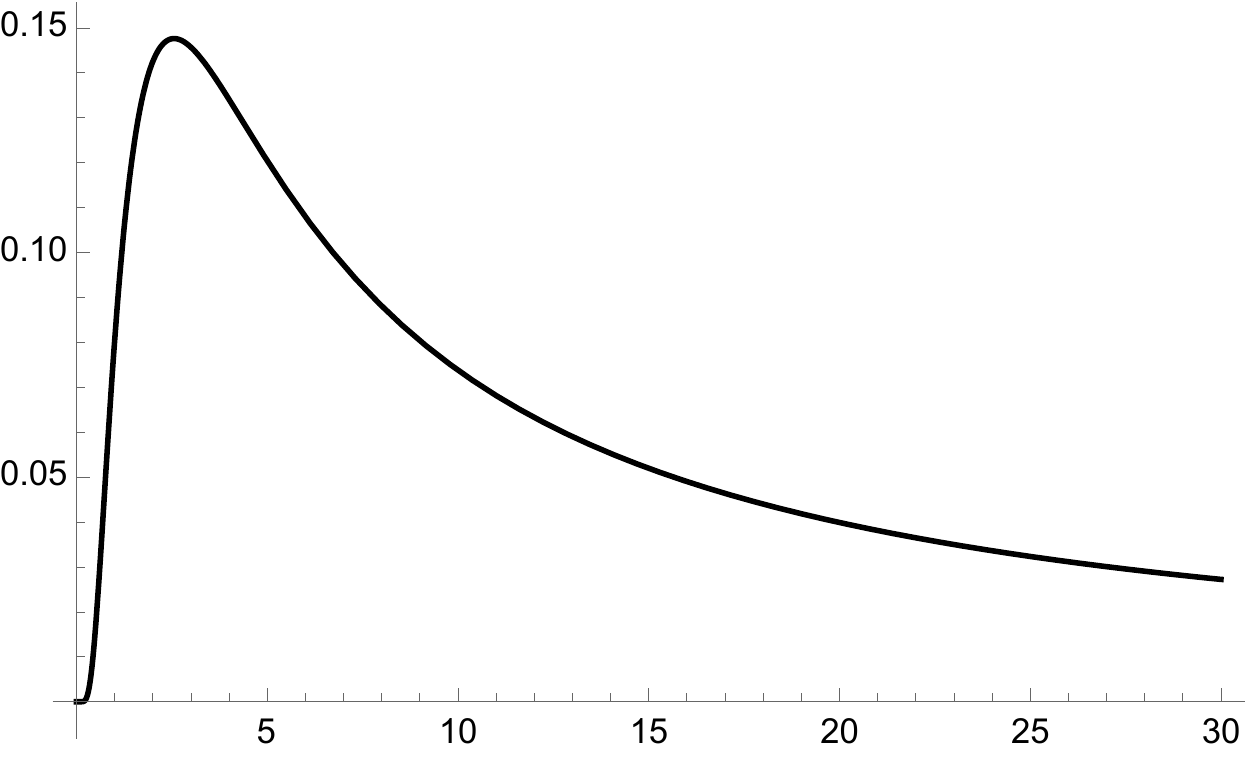}
\begin{picture}(0,0)
  \put(0,5){$H$}
   \put(-180,5){$H$}
   \put(-338,99){$\rho^{1\over 4}$}
   \put(-159,99){$\Delta(\rho^{1\over 4})$}
   \end{picture}
  \caption{One the left: The dependence $\rho^{1\over 4}(H)$ at fixed value of the mass
    $m=2$. The black curve is the expression \eqref{rhofinal4D},
    whereas the blue curve the flat space result
    \eqref{rhofreegas}. On the right: The difference of the two curves.}
  \label{fig:rhoH}
\end{figure}

However, like for a free gas in flat space with
\eq{
  \label{rhofreegas}
  \rho=\int {d^3k\over (2\pi)^3}  {w\over  e^{2\pi\omega/H}-1}=
  \int_m^\infty  {d\omega\over 2\pi^2} \sqrt{\omega^2-m^2}
   {w^2\over  e^{2\pi\omega/H}-1}
}
where $w^2=\vec k^2+m^2$,  for $H\ll m$  the
energy density is exponentially suppressed.
This can be seen in figures \ref{fig:rhom} and \ref{fig:rhoH}, where
we plotted the energy density against the mass at fixed value of
$H$ and against the temperature $H$ at fixed value of $m$. 

\vspace{0.3cm}
%%%%%%%%%%%%
%%%%%%%%%%%%
%%%%%%%%%%%%

\vspace{0.1cm}
%%%%%%%%%%%%
%%%%%%%%%%%%
%%%%%%%%%%%%

\noindent
Apparently, in the intermediate regime $m\sim H$
or equivalently $\lambda_c\sim \ell_{\rm dS}$,
the two curves 
differ and show kind of a resonant behavior,
while in the limiting low and high temperature regimes they
essentially agree. 
Note that the main
difference between the two expressions for $\rho$ is
that, opposed to the free gas result \eqref{rhofreegas},
in  \eqref{rhofinal4D} $\omega$ takes values in the full interval
$0\le \omega <\infty$. 

In case we now have a tower of massive states with masses $m_i$, like in string theory,
it is clear that the main contribution to the energy density (and also
the pressure) at a
fixed temperature $T=H/(2\pi)$ comes
from those states with masses below $T$, i.e. $m_i\ll H$.
As we have seen the contribution from states with masses  $m_i\gg H$
is exponentially suppressed. Moreover, the scaling of $\rho$ with
$H$ can already be determined from the flat space thermal expression
for the energy density \eqref{rhofreegas}. Recall from the previous
section that it is this scaling that  determines the scaling of the
quantum break time.

Even though we do not know how all the steps of this computation can be
generalized to string theory,
the simple form of the flat space contribution allows for a well-
motivated guess.
\begin{quotation}
  \noindent
{\it Assumption 3: Also in string theory on dS space there will
be a flat space thermal contribution to the energy density and
pressure that can be evaluated from the well known
stringy expressions for the perturbative thermal one-loop partition 
function and the resulting free energy.}
\end{quotation}

\noindent
We will make this more concrete in the following section.

\section{Thermodynamics of string theory}
\label{sec_3}

Motivated by the discussion in the previous section, 
we now consider the thermodynamics of strings at finite
temperature in  flat space. Here the story is a bit more intricate than in field
theories, mainly due to the existence of thermal winding modes
and the consistency condition of modular invariance of the 
thermal partition function.

\subsection{Free energy of strings at finite temperature}

To make this paper self-contained, let us recall in this section some salient features of the description
of strings at finite temperature (see e.g. \cite{Rohm:1983aq, Alvarez:1986sj, Atick:1988si, Dienes:2012dc}), where the main focus is on the
string theory generalization of the free energy and the energy
density.

First, let us consider a boson of mass $m$ at fixed temperature $T$ in
an $n$-dimensional space-time.
In this case, the free energy density ${\cal F}=F/V$  is defined in the usual way
as
%\footnote{Note that  via $\rho={\cal F}-T{\partial {\cal F}\over \partial T}$ this free energy indeed gives the
%expression \eqref{massive_rho_ndim_final}.}
\eq{
                {\cal F}_b(T)=-T\log Z_b(T)\,,\qquad {\rm where}\quad
                Z_b(T)=\prod_p {1\over 1-e^{-{E\over T}}}
}
with the dispersion relation $E^2=p^2+m^2$. Employing the relation
\eq{
         \log  (1-e^{-{E\over T}})={1\over 2}\sum_{m\in \mathbb Z}
         \log(E^2+4\pi^2 m^2 T^2) \, ,
}
one can express the free energy as 
\eq{
             {\cal F}_b(T)={T\over 2}\int {d^{n-1} p\over (2\pi)^{n-1}}
             \sum_{m\in\mathbb Z} \log\big(E^2+4\pi^2 m^2 T^2\big)\, ,
}
involving the infinite sum over the so-called Matsubara modes $m$. 
For a fermion, the analogous expression involves a sum over
half-integer Matsubara modes
\eq{
             {\cal F}_f(T)=-{T\over 2}\int {d^{n-1} p\over (2\pi)^{n-1}}
             \sum_{m\in\mathbb Z} \log\big(E^2+4\pi^2 \big(m+{\textstyle{1\over 2}}\big)^2 T^2\big)\,.
}
Identifying $R=1/(2\pi T)$, one can interpret the Matsubara modes as
Kaluza-Klein modes of a circle compactification, whose contribution
to the energy is as usual $E^2_{\rm KK}=m^2/R^2$. Therefore, the free energy of a gas of particles
at temperature $T$ in $d=n-1$ (spatial)  dimensions takes the same
form   as the  $n$-dimensional vacuum energy compactified on a
Wick-rotated time direction with radius $R=1/(2\pi T)$.

This picture can be generalized to string theory fairly
straightforwardly.
 Employing the usual steps for computing string
partition functions, one arrives at the following general
expression for the free energy of a string at temperature $T$
moving in $n$ uncompactified (flat) dimensions (in string frame)
\eq{
\label{freeenergy}
     {\cal F}(T)=-{T\over 2} \left({M_s\over 2\pi}\right)^{n-1} \int_{\cal
       F} {d^2\tau\over \tau_2^2} {1\over \tau_2^{{n\over 2}-1}} Z_{\rm
       string}(\tau,\ov\tau;T)\,,
} 
where the integral over the continuous momenta $p_i$, $i=1,\ldots,n-1$
has already been performed and we absorbed a factor of $\sqrt{\tau_2}$
into $Z_{\rm string}$ (see eq.\eqref{circlat} below). 
Hence, the string partition function
$Z_{\rm string}$ is over the remaining string modes, i.e. the string oscillator modes,
internal KK and winding modes and the Matsubara modes for the 
thermal circle compactification. Concerning the latter, one has to be a bit more
careful, as in string theory a circle compactification not only leads
to KK modes but also to winding modes. Moreover, one has to make 
sure that (space-time) bosons couple to integer and fermions to half-integer 
Matsubara modes in such a way that the final expression still features
modular invariance.

The resolution to all these issues can be best described in terms of a
string orbifold construction. The thermal partition function is closely
related to a winding Scherk-Schwarz orbifold (WSS), i.e. a string
theory compactification on $S^1/(-1)^F S_w$ with radius $R=1/(2\pi T)$.
Here $F$ denotes the space-time fermion number and $S_w$ the winding
shift that acts as $S_w:|m,n\rangle\to (-1)^n |m,n\rangle$ on a
KK/winding mode. To describe the general form of the orbifold partition
function let us introduce the following lattice contribution
\eq{
\label{circlat}
      Z_{\rm circ}(m,n)=\sqrt{\tau_2} \sum_{m,n}
      q^{{\alpha'\over 4}({m\over R} +{nR\over \alpha'})^2} \; \ov q^{{\alpha'\over 4}({m\over R} -{nR\over \alpha'})^2}\, ,
}
where $(m,n)$ indicate over which range the KK/winding modes run.
Denoting  
\eq{      
\label{latticesums}
           {\cal E}_0&=Z_{\rm circ}(m,2n)\,, \qquad\qquad {\cal E}_{1/2}=Z_{\rm
             circ}(m+{\textstyle{1\over 2}},2n)\,,\\
          {\cal O}_0&=Z_{\rm circ}(m,2n+1)\,, \qquad {\cal O}_{1/2}=Z_{\rm
             circ}(m+{\textstyle{1\over 2}},2n+1)\,,
}
the orbifold partition function in the untwisted sector can be
generally expressed as
\eq{
              Z^{\rm (WSS)}_{\rm u}(R)=Z_B\, {\cal E}_0(R) - Z_F\,  {\cal O}_0(R) \, ,
}
where $Z_{B/F}$ are the space-time boson/fermion contributions of the theory before
the orbifold. Applying  modular $T$ and $S$ transformations a twisted
sector will appear which can generally be expressed as
\eq{
         Z^{\rm (WSS)}_{\rm t}(R)=
              Z^{(1)}_t\, {\cal E}_{1/2}(R)  + Z_t^{(2)}\,  {\cal O}_{1/2}(R)\,,
}
where the form of $ Z^{(1,2)}_t$ depends on the theory in question. 
The sum of the untwisted and the twisted partition functions define
the modular invariant partition function  $Z^{\rm (WSS)}$
of the WSS orbifold, where the temperature dependence solely resides
in the lattice sums \eqref{latticesums}.

This is not yet the thermal partition function, but is closely related.
For the final step, let us consider the behavior of
\eqref{latticesums} under modular transformations. 
Under a modular T-transformation $ {\cal E}_{0}, {\cal E}_{1/2}, {\cal
  O}_{0}$ are invariant whereas $ {\cal O}_{1/2}$ receives a minus
sign. The action of a modular S-transformation is given by
\eq{
       S={1\over 2}\left(  \begin{matrix} 1 & 1 & 1 &1 \\
                                                          1 & 1 & -1 &
                                                          -1  \\  1 &
                                                          -1 & 1 & -1
                                                          \\ 1 & -1 & -1 &1   \end{matrix}  \right)\,.
}
Therefore, the ``thermal map''
\eq{
\label{thermalmap}
 {\cal T}:  \begin{cases} {\cal E}_{0} &\longleftrightarrow  {\cal
     E}_{0} \\
{\cal E}_{1/2} &\longleftrightarrow  {\cal
     O}_{0} \\
{\cal O}_{1/2} &\longleftrightarrow  {\cal
     O}_{1/2} \end{cases}
}
is an isomorphism and maps a modular invariant partition function
to another modular invariant partition function. Applying this
thermal map to the WSS orbifold, one gets the thermal partition
function
\eq{
\label{thermalP}
               Z^{\rm (T)}(T)=Z_B\, {\cal E}_0 (T)  - Z_F\,  {\cal E}_{1/2}(T)
                + Z^{(1)}_t\, {\cal O}_{0}(T)  + Z_t^{(2)}\,  {\cal O}_{1/2}(T)\,.
}
Note that now the untwisted space-time fermions indeed couple to  half-integer
KK Matsubara modes and we have used $R=1/(2\pi T)$.

\subsection{Examples: type IIB and heterotic}

Let us consider two simple ten-dimensional examples.
First, recall that the type IIB partition function can be expressed as
\eq{
                 Z_{\rm IIB}={1\over \eta^{8}(\tau) \eta^{8}(\ov\tau)   } \Big(\chi_V^{SO(8)}(\tau) -
               \chi_S^{SO(8)}(\tau) \Big) \Big(\ov\chi_V^{SO(8)}(\ov\tau) -
               \ov\chi_S^{SO(8)}(\ov\tau) \Big) \, ,
}
where the notation of the characters of the $\widehat{SO}(8)_1$ representations
is standard.
Compactifying the 10D type IIB  theory  on a circle and performing the winding Scherk-Schwarz
orbifold leads to the partition function
\eq{
      Z_{\rm IIB}^{\rm (WSS)}(R)={1\over |\eta|^{16}   }
      \Big(  &(\chi_V \ov\chi_V +\chi_S\ov\chi_S ){\cal E}_0 
- (\chi_V \ov\chi_S + \chi_S \ov\chi_V ){\cal O}_{0}\\
     +& (\chi_O \ov\chi_O +\chi_C\ov\chi_C ){\cal E}_{1/2} -
      (\chi_O \ov\chi_C +\chi_C\ov\chi_O ){\cal O}_{1/2}\Big)\,.
}
This partition function interpolates between the type IIB superstring ($R\to
0$)
and the non-supersymmetric tachyonic type 0B superstring ($R\to \infty$).
Applying the thermal map \eqref{thermalmap} leads to the  thermal partition function 
\eq{
\label{thermalIIB}
      Z_{\rm IIB}^{\rm (T)}(T)={1\over |\eta|^{16}   }
      \Big(  &(\chi_V \ov\chi_V +\chi_S\ov\chi_S ){\cal E}_0 
- (\chi_V \ov\chi_S + \chi_S \ov\chi_V ){\cal E}_{1/2}\\
     +& (\chi_O \ov\chi_O +\chi_C\ov\chi_C ){\cal O}_{0} -
      (\chi_O \ov\chi_C +\chi_C\ov\chi_O ){\cal O}_{1/2}\Big)\,.
}
One observes that the third term in \eqref{thermalIIB} develops a tachyonic
winding mode $(m,n)=(0, \pm 1)$ for temperatures larger than
\eq{
\label{tacwindow}
                  T > {1\over \sqrt{2}}{M_s\over 2\pi}\,.
}
This is nothing else than the Hagedorn temperature $T_H$, which generically
is supposed to appear in theories with an exponential growth 
of degrees of freedom. However, the partition function reveals
that it is an infrared effect associated with a massless (tachyonic)
mode developing at $T_H$.

As a second  example, we consider the  ten-dimensional heterotic string
with the partition function
\eq{
               Z={\Lambda^{G}(\ov\tau)\over \eta^{8}(\tau) \eta^{8}(\ov\tau)   } \Big(\chi_V^{SO(8)}(\tau) -
               \chi_S^{SO(8)}(\tau) \Big) .
}
Here the characters $\chi^{SO(8)}_V(\tau)$ and $\chi^{SO(8)}_S(\tau)$ arise from the left-moving
fermions and $\Lambda^{G}(\ov\tau)$ denotes the character of the
$16$ right-moving bosons compactified on the root-lattice of
$SO(32)$ or $E_8\times E_8$, respectively.
Compactifying the 10D heterotic theory  on a circle and performing the winding Scherk-Schwarz
orbifold and applying the thermal map leads to the partition function
\eq{
\label{hetthermal}
      Z^{\rm (T)}(T)={\Lambda^{G}\over |\eta|^{16}   }
      \Big(\chi_V^{SO(8)} {\cal E}_0 -
               \chi_S^{SO(8)} {\cal E}_{1/2}- \chi_C^{SO(8)}  {\cal O}_{0}
     +\chi_O^{SO(8)} {\cal O}_{1/2}\Big) \,.
}
Note that in this case, on the level of characters one has $ \chi_S^{SO(8)}=
\chi_C^{SO(8)}$ and therefore $Z_F=Z_t^{(1)}$. Moreover, let us mention that the last term in
\eqref{hetthermal} develops a tachyon in the window 
\eq{
\label{tacwindow2}
                  (2-\sqrt{2}) {M_s\over 2\pi} < T < (2+\sqrt{2})
                  {M_s\over 2\pi}\,.
}
Concretely, it is the KK/winding modes $\pm (-1/2,1)$ from ${\cal
  O}_{1/2}$  that can become tachyonic.

\subsection{Tachyon condensation and a new phase of strings}
\label{sec_tac}

We have seen in these two examples that at a certain critical temperature thermal tachyons appear
in the spectrum. 
As argued in the seminal work by
Atick-Witten \cite{Atick:1988si}, this rather indicates a phase transition (where
the winding mode gets a non-zero vacuum expectation value) than a
fundamental maximal temperature for strings. In fact, it is argued
that this is a first order phase transition. 
Considering the effective theory of this condensing mode, they argue that 
also a tree-level (genus 0) contribution to the free energy is
generated that  strongly backreacts on the geometry. Therefore, 
analogous to the deconfining phase transition in QCD, 
this phase transition is expected to be dramatic and will reveal 
new degrees of freedom. As far as we are aware, there is no consensus
yet what these new degrees of freedom of string theory at very high
temperature (energy) are and by which theory they are described.
It might be related to black-holes and/or a  topological gravity theory.

Even though neither the meaning
of the canonical ensemble  nor the finiteness of the partition
function is clear in this new phase, Atick-Witten argue 
that the extrapolation of the 
one-loop free energy \eqref{freeenergy} nevertheless gives a reasonable result for the high
temperature dependence. 
For the type IIB superstring, this deviates from the usual field theory
result and scales with $T$ like
\eq{
\label{atickwitten}
                    {\cal F}(T)\sim   \Lambda^{(1)}_{\rm 0B}\;  T^2 \, ,
}
where $\Lambda^{(1)}_{\rm 0B}$ denotes the (diverging) one-loop cosmological constant of the
type 0B theory. 
For the heterotic string, it is the lower critical temperature 
$T_H=(2-\sqrt{2}) {M_s\over 2\pi}$, at  which a phase transition is
expected to occur. However, at least formally the  model is 
tachyon-free above a second critical temperature\footnote{In \cite{Angelantonj:2008fz}
a thermal partition function for a certain type IIB
asymmetric orbifold was constructed that was free of  tachyons for 
any temperature.}, which puts some
more confidence on the  scaling
\eq{
\label{atickwitten_het}
                    {\cal F}(T)\sim   \Lambda^{(1)}_{\rm het}\;  T^2
}
in the high temperature regime.

Let us mention that the quadratic temperature dependence 
could  be explained by generalizing T-duality to the
thermal circle  (see also \cite{Dienes:2003sq,Dienes:2003dv}).
Recall that  a pure circle compactification of string theory enjoys 
T-duality acting as $R\to \alpha'/R$, which will map for instance
the two following Scherk-Schwarz orbifolds to one another \cite{Angelantonj:2002ct}: type IIB on
$S^1/(-1)^{F} S_w \longleftrightarrow$ type IIA on $S^1/(-1)^{F} S$,
where $S$ denotes the left-right symmetric momentum shift. 
However,  inspection reveals that there exists the slightly modified transformation $R\to
\alpha'/2R$,  under which  the lattice sums \eqref{latticesums} transform as 
\eq{
     \{ {\cal E}_0, {\cal E}_{1/2}, {\cal O}_0,  {\cal O}_{1/2}\}\to
      \{ {\cal E}_0, {\cal O}_{0}, {\cal E}_{1/2},  {\cal O}_{1/2}\}\, ,
}
which is nothing else than the thermal map. Taking into account that 
this transformation  acts on the temperature as
\eq{
\label{tempdual}
                T\to {T_c^2/ T}\,,\qquad {\rm with}\quad T_c=\sqrt{2}
                {M_s\over 2\pi}\,,
}
we find the following transformation for the partition functions
\eq{
                     Z^{\rm (T)}(T_c^2/T)=Z^{\rm (WSS)}(T)\,.
}
This again shows the intricate relationship between these two
partition functions.  
For the free energy this implies
\eq{
                     {\cal F}^{\rm (T)}(T^2_c/T)=\left({T_c\over T}\right)^2 {\cal F}^{\rm (WSS)}(T)\,.
}
Applying this for instance to the type IIB example, one obtains
\eq{
        \lim_{T\to \infty}  {\cal F}^{\rm (T)}(T)\sim T^2 \lim_{R\to \infty}
      {\cal F}^{\rm (WSS)}(R) \sim T^2  \Lambda^{(1)}_{\rm 0B}\,,
}
which shows that the quadratic temperature dependence is a universal
scaling relation.
The energy density  and the pressure are given in terms of the free energy density as
\eq{
\label{energydensity}
  \rho={\cal F}-T{\partial {\cal F}\over \partial T}\,,\qquad p=-{\cal F}\, ,
}
from which one can derive
\eq{
\label{dualityenergy}
                 \rho^{\rm (T)}(T_c^2/T)=-\left({T_c\over T}\right)^2 \rho^{\rm (WSS)}(T)\,.
}
Note that both $T_c$ and $T_H$ are of the order
of the string scale but differ by numerical prefactors.

To summarize: In the ultra high temperature region $T>T_H$ 
the condensed phase of strings is argued to have fewer degrees of
freedom  than naively expected and shows a quadratic scaling  $\rho\sim T^2$
of the energy density with temperature. 
Support for this claim comes from a naive  extrapolation of the
perturbative expressions for the free energy into the regime $T>T_H$ 
and from a T-duality argument.
Thus, quantum gravity at high energies behaves rather as a two-dimensional field
theory. Intriguingly, the same observation was made in other
approaches to quantum gravity, like 
the  causal dynamical triangulation or asymptotic safety
programs (see e.g.\cite{Carlip:2009kf,Carlip:2017eud}).
In the following we will assume that this   estimate is
indeed correct and investigate what the consequences 
for quantum breaking and the swampland conjectures are.
It is clearly  very important to support the assumed scaling  by, for
instance, a direct computation for  a candidate of a topological
gravity theory or any other  theory suggested to be valid beyond the 
Hagedorn transition.

\section{dS quantum breaking for strings}
\label{sec_4}

Assume one has constructed an $n$-dimensional tree-level string model that features
moduli stabilization in a tachyon-free dS or at least quasi dS phase.
As we have seen in section \ref{sec_2}, from the semiclassical approach of quantizing a free boson
in the static patch, one obtains a matter contribution to the energy density and pressure.
It is thermal with the Gibbons-Hawking temperature $T=H/(2\pi)$.
Including this contribution in the Einstein equation,  i.e.  taking the quantum backreaction of the
scalar into account, one finds that dS can no longer be a solution.  Interpreting this as the source of 
quantum breaking of de Sitter,  we aim to directly generalize this to string theory.

As was shown in section \ref{sec_emtensor},  for massive scalars,  such as the ones in
stringy towers of states, there is significant coincidence between the flat space result and the actual values of $\rho, p$ in static dS.  More importantly,  the scaling of these quantities with the temperature, which dictates the quantum breaking time, can be read off the flat space energy density.
Hence, for our desired generalization to string theory,  it seems sufficient to invoke the usual methods for 
the computation of the thermal partition function,  which in turn directly leads to the energy and pressure temperature dependence.

Of course, for a concrete flux compactification with moduli
stabilization the full one-loop thermal partition function cannot
be determined explicitly. The models are usually described only
in an effective supergravity approach, where only the lightest states
are kept. Fortunately, the low ($T< T_H$) and the (naively extrapolated) high
($T> T_H$) temperature behavior turn out to be quite universal and
do not depend on the full string theory spectrum.
Let us discuss this in more detail.

\subsection{Low temperature regime}

Consider first the low-temperature regime and assume that
in this regime there are no tachyons in the spectrum. 
In this case, all the winding modes with $n\ne 0$ are very massive
and give a  highly suppressed contribution to  the free energy \eqref{freeenergy}.
Therefore, only the KK modes in ${\cal E}_0$ and ${\cal E}_{1/2}$
need to be considered which  couple to $Z_B$ and $Z_F$ in
\eqref{thermalP}.
First, one can show that in the $T\to 0$
limit, the free energy approaches the 1-loop vacuum energy
\eq{
               \lim_{T\to 0} {\cal F}(T)=\Lambda^{(1)}_0\,.
}
For finite $T$ the general behavior of the free energy is already
revealed
by considering the   lightest states of mass $M_b$ and $M_f$ in  $Z_B$ and $Z_F$.
Then, the main contribution to the free energy is expected to arise from
the following terms
\eq{
\label{lowTF}
     {\cal F}(T)\sim -{T\over 2} &\left({M_s\over 2\pi}\right)^{n-1} \!\!\int_{1}^\infty
       {d\tau_2\over \tau_2^{n+1\over 2}}  \biggl[ e^{-\pi\tau_2 \big({M_b\over
         M_s}\big)^2}\, \sum_{m=-\infty}^{\infty}
       \exp\left(-2\pi\tau_2 m^2 ({\textstyle{T\over T_c}})^2\right)\\[0.1cm]
      &\phantom{aaaaaaaa}-e^{-\pi\tau_2 \big({M_f\over
         M_s}\big)^2}\, \sum_{m=-\infty}^{\infty}
       \exp\left(-2\pi\tau_2 (m+{\textstyle{1\over 2}})^2 ({\textstyle{T\over T_c}})^2\right)\biggr]\,
}
where we have carried out the integration over $\tau_1$ that gives the
level-matching constraint. Note that here we are using the
``self-dual'' temperature $T_c$, which differs from the Hagedorn
temperature only by an order one numerical prefactor. 
Using \eqref{lowTF}  one can straightforwardly compute the energy density $\rho$ via
\eqref{energydensity} and evaluate it numerically.

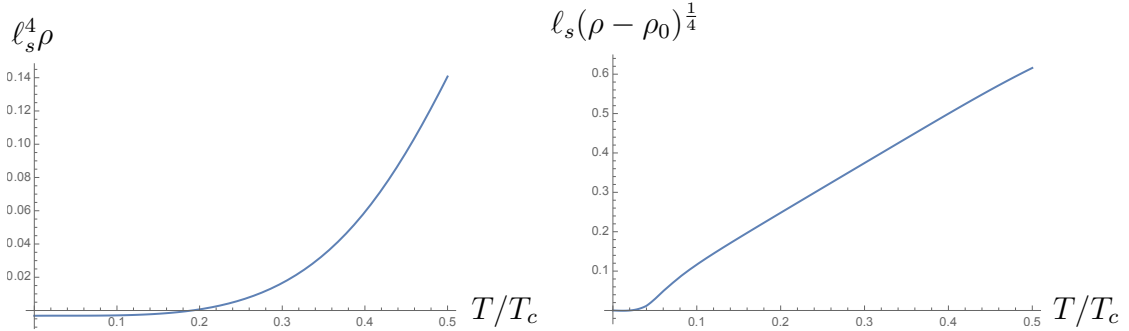
\begin{figure}[ht]
\vspace{0.4cm}
  \centering
   \includegraphics[width=6cm]{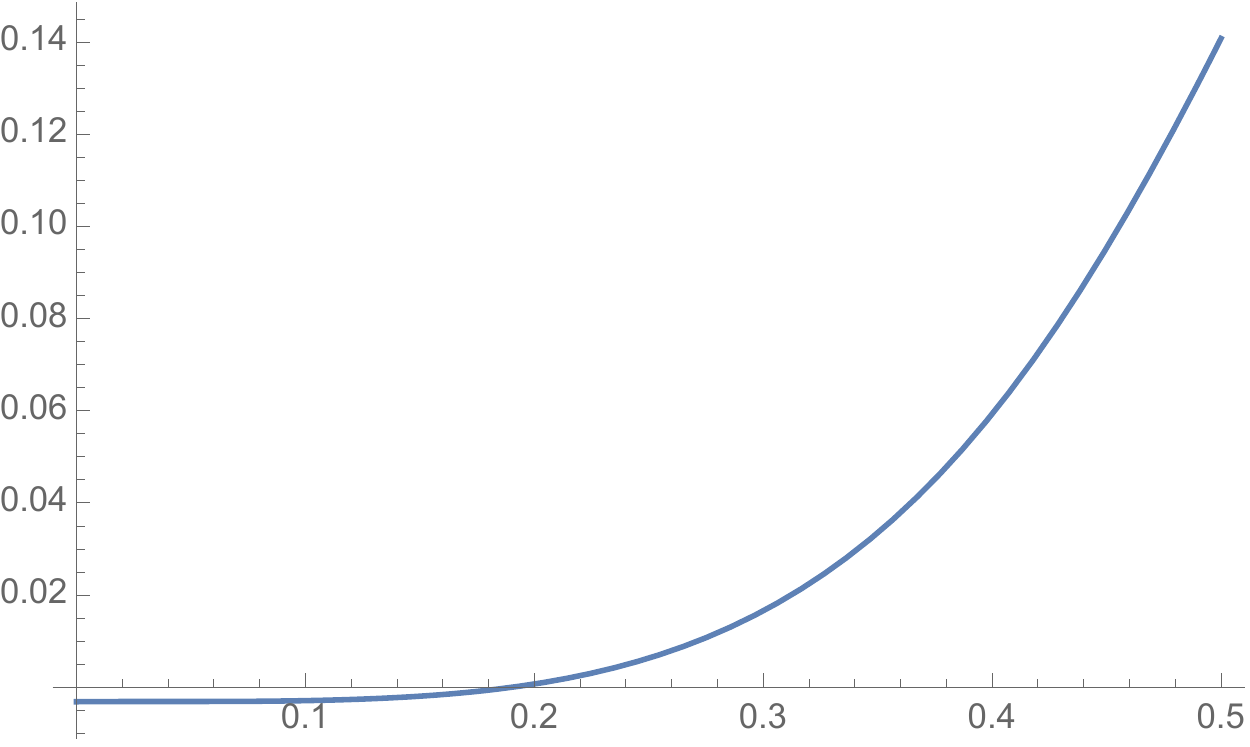}
\hspace{1.5cm}
   \includegraphics[width=6cm]{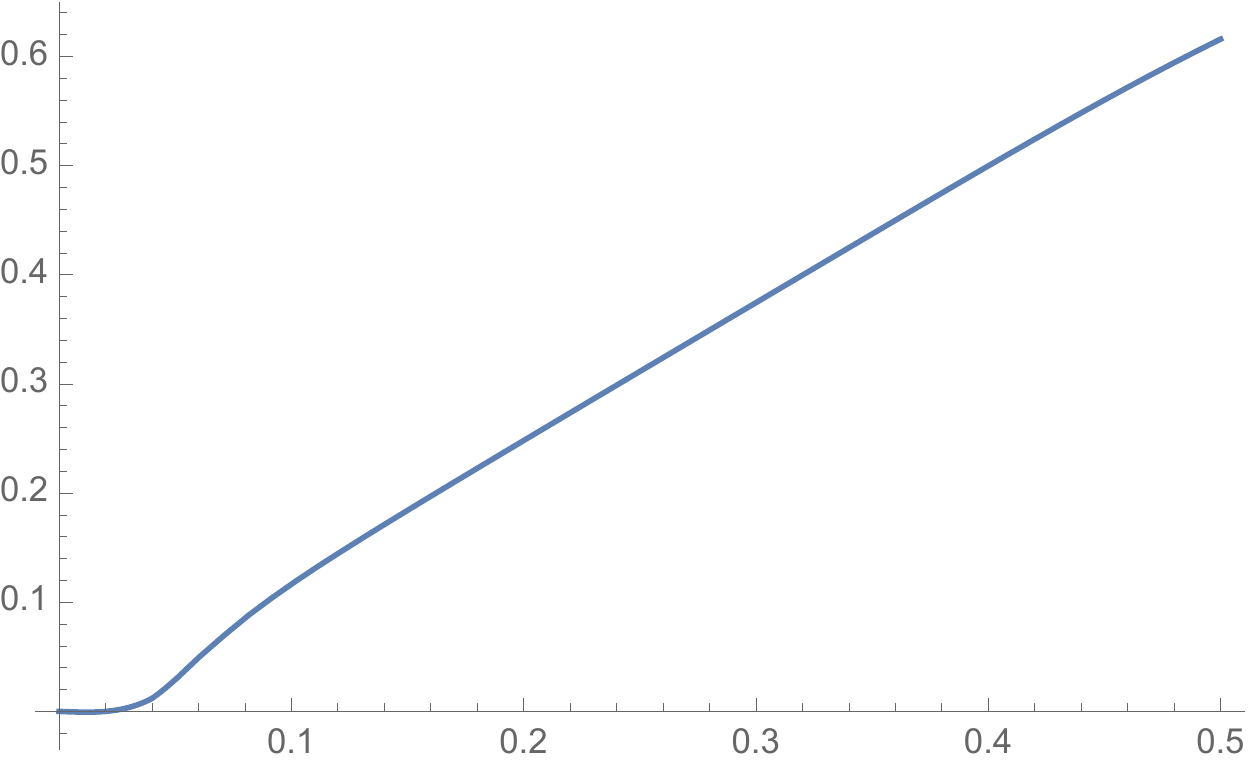}
\begin{picture}(0,0)
    \put(0,5){$T/T_c$}
    \put(-220,5){$T/T_c$}
    \put(-190,113){$\ell_s (\rho-\rho_0)^{1\over 4}$}
   \put(-394,109){$\ell_s^4\rho$}
    \end{picture}
  \caption{Left: The 1-loop energy density in the low temperature regime
    $T<T_c$ for $n=4$. Right: The fourth root of $\rho-\rho_0$. 
   Choice of parameters: $m_B=0$, $m_F=0.2$.}
  \label{fig:1}
\end{figure}

In figure \ref{fig:1} we show the resulting behavior of $\rho$ in the
low-temperature regime in four dimensions. The right-hand side 
reveals that the temperature dependence for $T^2>{\rm Str}(M^2)$ is indeed
quartic, as expected from the Stefan-Boltzmann law. 
Performing an analogous computation in arbitrary dimension $n$, one
can confirm  the scaling  ${\cal F}\sim T^n$. For even lower
temperatures one
finds the usual exponential drop-off  $\exp(-m/T)$ expected from field
theory for $T<m$.  
%%%%%%%%%%%%
%%%%%%%%%%%%
%%%%%%%%%%%%
Moreover, one can numerically check that the equation of state in the 
regime ${\rm Str}(M^2)<T^2<T^2_c$ is indeed $p=\rho/(n-1)$.

To summarize, in the low temperature regime ${\rm Str}(M^2)<T^2<T^2_c$ the string
theoretic expression for the 
free energy nicely reproduces  the expected low-energy field theory
results and in particular the $\rho\sim T^n$ scaling.
%Concerning the thermal matter arising from a coarse-grained
%energy-momentum tensor in the Bunch-Davies vacuum, in this temperature
%regime string theory does not deviate from an effective  purely field theoretic treatment.

\subsection{High temperature regime}

Increasing the temperature, we have already seen for the
example of the type IIB superstring  that one
encounters a critical temperature $T_H$, where the appearance of thermal tachyons
signals a phase transition of the system. It is clear that such
tachyons
can only arise from the terms  ${\cal O}_0$ and ${\cal O}_{1/2}$,
that are multiplied in  \eqref{thermalP} by the twisted sectors $Z_t^{(1,2)}$. Therefore, 
thermal tachyons can only appear when $Z_t^{(1,2)}$ themselves
contain tachyons. The general expectation from the exponential
growth of string states is that this is a generic phenomenon. However, 
the precise value of the temperature where this happens might be 
a model dependent question. 

Following Atick-Witten \cite{Atick:1988si}, we now assume
that the temperature dependence of the free energy in this new phase
of string theory can still be reliably estimated from the expression 
\eqref{freeenergy}, while ignoring the tachyonic divergences.
In this high $T$ regime, the KK modes with $m\ne 0$ give a negligible
contribution to the free energy so that solely the winding  modes in ${\cal E}_0$ and ${\cal O}_{0}$
need to be considered. These couple to $Z_B$ and the twisted sector
states $Z^{(1)}_t$ in \eqref{thermalP}.
The lightest state in $Z_B$ will be  a boson of mass $M_b$, while
the lightest state in  $Z^{(1)}_t$ could be a fermion or a
boson of mass $M_t$.  Thus, the main contribution to the free energy will come from 
the following terms
\eq{
\label{freeenergyhigh}
     {\cal F}(T)\sim -{T\over 2} &\left({M_s\over 2\pi}\right)^{n-1} \!\!\int_{1}^\infty
       {d\tau_2\over \tau_2^{n+1\over 2}}  \biggl[ e^{-\pi\tau_2 \big({M_b\over
         M_s}\big)^2}\, \sum_{n=-\infty}^{\infty}
       \exp\left(-2\pi\tau_2 n^2 ({\textstyle{T_C\over T}})^2\right)\\[0.1cm]
      &\phantom{aaaaaaaa}\mp e^{-\pi\tau_2 \big({M_t\over
         M_s}\big)^2}\, \sum_{n=-\infty}^{\infty}
       \exp\left(-2\pi\tau_2 (n+{\textstyle{1\over 2}})^2 ({\textstyle{T_c\over T}})^2\right)\biggr]\,.
}

\begin{figure}[ht]
  \centering
  \includegraphics[width=8cm]{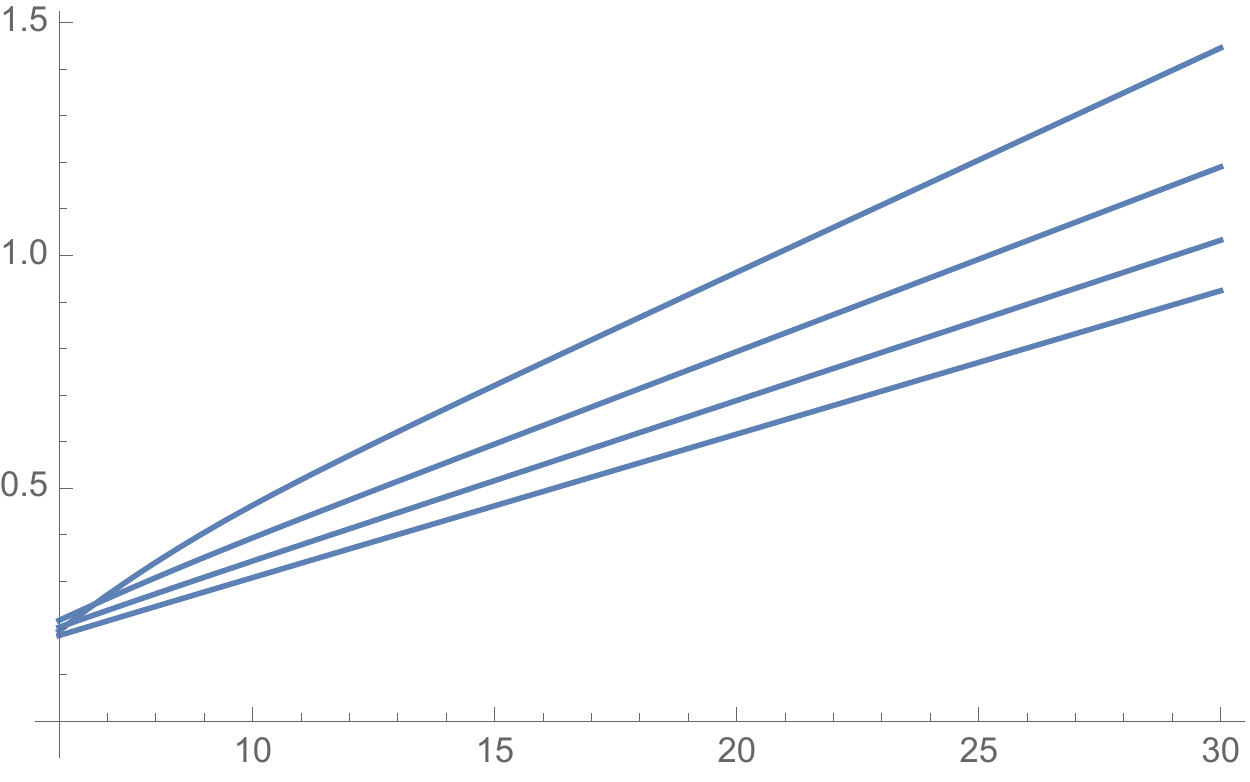}
\begin{picture}(0,0)
    \put(0,10){$T/T_c$}
    \put(-5,84){$n=7$}
    \put(-5,97){$n=6$}
     \put(-5,110){$n=5$}
     \put(-5,132){$n=4$}
   \put(-233,146){$(\ell_s^n\rho)^{1\over 2}$}
    \end{picture}
  \caption{The square root of the energy density in the high temperature regime
    in dimensions  $n=4,5,6,7$. 
   Choice of parameters: $m_B=0.1$, $m_F=0.2$.}
  \label{fig:5}
\end{figure}
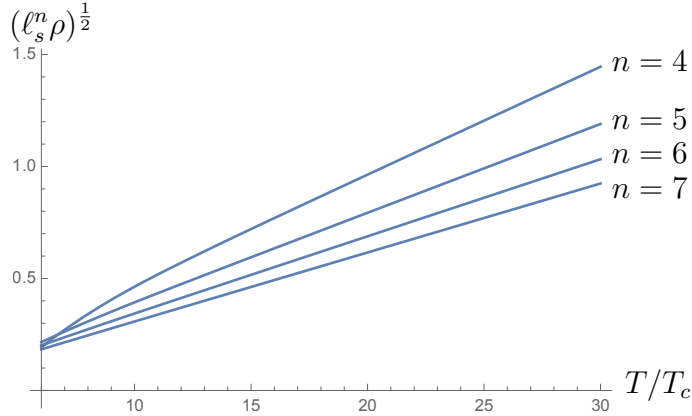

As  shown in figure \ref{fig:5},  
independent of the number of dimensions one finds a 
quadratic behavior $\rho(T)\sim T^2$.

\vspace{0.3cm}
%%%%%%%%%%%%
%%%%%%%%%%%%
%%%%%%%%%%%%

\noindent
This can be easily proven from \eqref{freeenergyhigh} by approximating the winding sum by a
Gaussian integral. However, it also follows from the duality relation \eqref{dualityenergy}
if in the limit $R\to \infty$ the WSS one-loop partition function approaches $C^{(1)}$, i.e.
\eq{
                 \rho(T)\to   -\left({T\over T_c}\right)^2 \lim_{R\to
                   \infty} \rho^{\rm (WSS)}(R)=-C^{(1)}\left({T\over T_c}\right)^2 \,.
}
Furthermore, the equation of state in this regime is
$p=\rho$. This suggests a drop in the number of degrees of freedom
in this regime and rather shows the scaling behavior of  a two-dimensional free gas.
Let us summarize the main findings:
\begin{itemize}

\item{In the low temperature regime, the string theoretic
    thermal energy density in $n$ dimensions behaves in the same way
    as in field theory, namely
   $\rho\sim T^{n}$ with equation of state $p={\rho\over n-1}$.} 
\item{In the high temperature regime, the scaling is universally
    $\rho\sim M^{n-2}_s  \,T^2$ with $p=\rho$, independent of the number
    of dimensions.}
\end{itemize}

Let us stress again, that this result was obtained by a  naive extrapolation of
the perturbative thermal string partition to a regime where it is 
not converging. 
The actual claim is  that in the new
 phase beyond the Hagedorn transition, carrying out 
a bona fide computation will also lead to the same quadratic scaling behavior.

\subsection{dS quantum breaking for strings}

Now let us come back to our initial question about the behavior of the
thermal quantum matter in a dS space.  As mentioned, we are going to assume that the
quantum backreaction approach again leads to a thermal quantum
matter component, whose energy density and pressure are given in terms
of the  thermal string partition function \eqref{freeenergy} with $T=H/(2\pi)$. 
Of course this is a strong assumption, whose proof is beyond our
computational abilities (at the moment), but it is a natural
extrapolation of the quantum field theory results. 

Up to this point, the computations in this section were done in the string frame.
In string theory, the Planck scale is a
derived quantity and related to the string scale via
\eq{
\label{stringplanckrel}
             M_{s}^{n-2}\sim { g_s^2 M_{\rm pl}^{n-2}\over  {\cal V}} \, ,
}
where  ${\cal V}$ denotes the volume of the internal $10-n$
dimensional space in units of the string scale and  $g_s$  the string coupling constant. 
Therefore,  in the perturbative
regime the string scale can be smaller than the Planck scale.
Thus, in the following we consider this case with $g_s\ll 1$ and ${\cal V}\gg 1$.

The large volume implies  that the free energy \eqref{freeenergyhigh}
will also contain a sum over  light Kaluza-Klein modes. For a single
circle direction and large radius $R$, this sum can be approximated by a Gaussian integral
as
\eq{
    \sum_{m\in \mathbb Z} e^{-\pi\tau_2 {m^2\over (M_s
        R)^2}}\approx    M_s R \int_{-\infty}^\infty dx \,e^{-\pi\tau_2
      x^2}={M_s R\over \tau_2^{1/2}}\,.
}
This shows the appearance of a multiplicative factor 
$M_sR\sim {\cal V}^{1/(10-n)}$  
and the expected increase of the power of $\tau_2$ (see
eq.\eqref{freeenergy}) in the decompactification limit.

Next we discuss the low and the high temperature phase separately,
where following \cite{Agrawal:2020xek} we  denote these two phases
as phase II and phase I, respectively.

\subsubsection*{Low $T$ regime (phase II)}

We have seen that for $T< T_H$ string theory nicely reproduces
the field theory result for the energy density and the pressure. 
Thus, we expect to just recover the results from section
\ref{sec_quantumbreak}, giving us the no eternal inflation constraints 
from the censorship of quantum breaking.

However, one needs to be a bit more careful.
 First,  in section
\ref{sec_quantumbreak} it was implicitly assumed that 
the equations of motion are just the Friedmann equations
resulting from the Einstein-Hilbert term.  It is known that 
string theory generates higher derivative  corrections to that.
These $\alpha'$ corrections involve higher curvature terms 
and for a dS background lead to $(H/M_s)^n$ corrections to the left-hand
side of the usual Friedmann equations. %\eqref{friedmaneq}. 
Therefore, at least at low temperatures, i.e. $H< M_s$, we are 
in the controlled regime.

Second, for ${\cal V}\gg 1$ but still working in an effective
$n$-dimensional theory, the just observed appearance of the factor
${\cal V}$ in the free energy gives rise to a slight  modification.
Hence, in this case one gets $\rho\sim  {\cal V} H^n$ such that the
quantum breaking time \eqref{quantumbreaktime} gets modified as
\eq{
              t_Q\sim   { M_{\rm pl}^{n-2}\over  {\cal V}
                H^{n-1}}\sim { M_{s}^{n-2}\over g_s^2  
                H^{n-1}}\,.
}
Censoring quantum breaking then leads to 
\eq{
                {|V'|\over V}\gtrsim g_s \left( {H\over
                    M_s}\right)^{n-2\over 2}\,,
}
where we have set $M_{\rm pl}=1$. 
Note that  the right-hand side still scales like $V^{n-2\over 4}$, as
it appears in the no eternal inflation bound, but in practice imposes a stronger bound than
\eqref{noeternalinflndim}. 
This implies that eternal inflation is censored by a string theoretic
quantum breaking bound.

The appearance of the string scale instead of the Planck scale
in $t_Q$ can be understood from the fact that for a large number 
of light species $N_{\rm sp}$  the effective
cut-off of quantum gravity is not the Planck scale but the species
scale \cite{Dvali:2007wp}
\eq{
\label{speciesscale}
\Lambda_{\rm sp}={M_{\rm pl} \over {N_{\rm sp}}^{\frac{1}{n-2}}}\,.
} 

To see what this means in our case,  say one has an effective theory in $n$ dimensions that 
has a tower of states with masses  $m_N=N \Delta m $,
with a degeneracy of states at each mass level that scales like $N^K$.  
As in \cite{Heidenreich:2017sim,Grimm:2018ohb,Heidenreich:2018kpg}, the number of species lighter than the species scale is given by
\eq{
              N_{\rm sp}=\sum_{N=1}^{\Lambda_{\rm sp}/\Delta m}  N^K
              \approx    \left( {\Lambda_{\rm sp}\over \Delta m} \right)^{K+1}\,.
}  
Consider compactifications on,  for instance,  isotropic $10-n$
dimensional tori or spheres.  In this
case one can simply deduce the degeneracy of KK modes as $K=9-n$
with a mass splitting $\Delta m=M_s/r$. Here $r$ denotes the radius
in units of the string length. Using as well the relation \eqref{stringplanckrel} with
${\cal V}=r^{10-n}$, one can solve the above the relations for the
species scale and get $\Lambda_{\rm sp}\sim M_s$. Therefore,  for
${\cal V}\gg 1$ a  tower of KK modes will  become light so that the
effective cut-off of quantum gravity 
becomes the string scale instead
of the Planck scale. This makes it less surprising that in the
relations above the string scale appears and not the Planck scale, as in
the simple $n$-dimensional field theory result where no large extra 
dimensions were included.

\subsubsection*{High $T$ regime (phase I)}

Next we  consider the  high temperature
regime $T> T_H$ with its proposed universal 2D-like scaling behavior.
This means that we study a dS space
with $H>M_s$, a regime that is usually not considered in any string
theory realization. In fact, one normally employs an effective field
theory description of string cosmology that is beyond control in this
regime, where $\alpha'$ corrections will be substantial. 

Indeed, the just discussed $H/M_s$ corrections to the left-hand
side of the Friedmann equations are now non-negligible
and it is a priori not clear how to extrapolate the latter beyond the
phase  transition\footnote{One could contemplate that the
T-duality \eqref{tempdual} extends also to the equation of motion  so that in the high $H$
phase one has an expansion in $M_s/H$.}. 
Remarkably,  after identifying the
temperature with the curvature of de Sitter space, also from this perspective
one is losing control at $H\sim M_s\sim T_H$.

On the contrary, the
methods we used for the description of the thermal  string theory
partition function or its relative, the WSS compactification, are of
CFT type and are capturing in principle all $\alpha'$ corrections (with
$g_s\ll 1$ still in the perturbative regime). In fact, the scaling
$ \rho \sim M^{n-2}_s \, T^2$ was very generic and not depending
on the many details of the remaining part of the partition function, i.e.
$Z_{B/F}$ and  $Z^{(1,2)}_{t}$. 

We now proceed under the working assumption  that in phase I, 
even though it is largely unknown, the two following aspects hold:
\begin{itemize}
\item{At leading order, the equation of motion is still given by 
    Einstein's equations, i.e. in this regime there exists  a controlled  gravity theory.}
\item{The energy density scales quadratically with the
    temperature.}
 \end{itemize}

Taking also the extra factor of ${\cal V}\sim (M_s R)^{10-n}$ into account,
in Einstein-frame the $n$-dimensional one-loop thermal energy density can be written as
\eq{
\label{scalerho}
      \rho=\kappa_n\, g_s^2\, M^{n-2}_{\rm pl} H^2 \,.
}
Now proceeding as in section \ref{sec_quantumbreak}, for this new (stringy) quantum
matter component with  equation of state parameter $w_m=1$, one
finds
%that the Hubble parameter evolves in time as
%\eq{
%   H(t)={H_0\over \left(1+{2\kappa_n g_s^2 \over (n-2)}  H_0 t\right)}\,.
%}
%This leads to
a decay of the dS vacuum with  a life-time
\eq{
               t_Q\sim  {1\over  g_s^2\, H}\,.
}
Note that this scales
  precisely as suggested in \cite{Dvali:2018fqu, Dvali:2018jhn}, namely $t_Q\sim 1/(\alpha H)$ with
$\alpha\sim g_s^2$. As mentioned in  the introduction, $\alpha$
denotes the  quantum interaction strength.
Censoring  quantum breaking and assuming that also in phase I
  there can be a scalar field with a scalar potential, one now gets for the slow-roll
parameter a lower bound
\eq{
           {|V'|\over V} \gtrsim c\,g_s\,.
}
Note that the right-hand side is independent of the potential $V$ in any space-time dimension,
but still scales like $g_s$ and hence  is not a pure $c$-number as in
the dS conjecture and the asymptotic limit of TCC. 

However, it might be that this
perturbative factor of $g_s$ is just an artefact of the extrapolation 
of the one-loop thermal partition function to the high temperature
phase that appears after tachyon condensation. 
Recalling from section \ref{sec_tac} that   tachyon condensation
induces  a tree-level contribution to the free energy,
one could very well imagine that  by directly performing the
computation  in this  mysterious phase of string theory, 
the factor of $g_s^2$ in \eqref{scalerho} will turn out to be  absent  such that $\alpha=\mathcal{O}(1)$.

%Similarly, for a tachyonic saddle point one naively finds the bound
%\eq{
%  {|V''|\over V} \gtrsim  c' \,g_s^2\,.
%}
%again as it appears in the refined dS swampland conjecture. 
%These bounds suggest that the two parameters in the refined
%dS swampland conjecture are related as $ c'\sim c^2$,
%at least as far as the  scaling with $n$ is
%concerned. Comparing this to the TCC, it is tempting
%to read off  from \eqref{tcc} the scaling $\kappa_n\sim {(n-2)/ (n-1)}$ and as a
%consequence predict  $c'\sim {1\over (n-1)(n-2)}$. 
%This indeed agrees with the numerical
%prefactor appearing in the bound  for ${|V''|/V}$ derived from the TCC \eqref{TCCtachy}.

\subsubsection*{Concluding remarks}

\begin{figure}[htb]
  \centering
   \includegraphics[width=9.5cm]{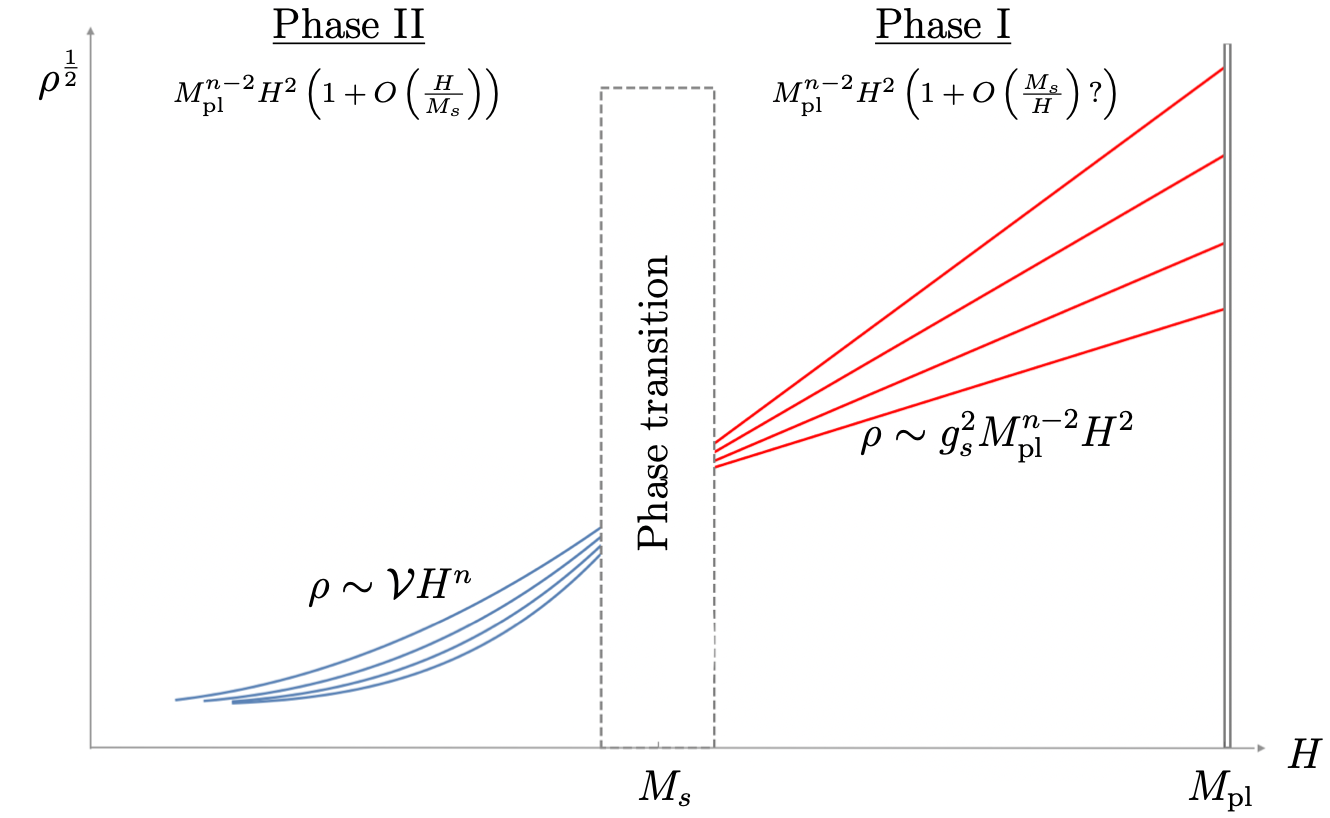}
     \caption{Schematic behavior of the square root of the energy density  in the perturbative case with
     $M_s\ll M_{\rm pl}$ for dimensions $n=4,5,6,7$. }
  \label{fig:summary1}
\end{figure}

Let us  schematically summarize the main results about quantum
breaking in string theory. In figure \ref{fig:summary1} we display
the left-hand side of the Friedmann equations
and the scaling of the energy density $\rho$  in the low and high temperature phases
of string theory.

The energy density changes from $n$-dependent at
low temperatures (phase II) to the universal $\sim H^2$ behavior at high
temperatures (phase I). Moreover, in phase II the left-hand side of the Friedmann equations
is under control and at leading order just given by the term following 
from Einstein's equations,  whereas in phase I we were just assuming
that a similar story holds.
Recall  that  in phase II we could employ computational  methods from perturbative 
string theory to really derive the quantities in question, whereas for 
the high energy phase I we had to rely on reasonable assumptions
that still need to be confirmed by a bona fide computation in e.g.
a topological gravity theory.

\begin{figure}[htb]
  \centering
   \includegraphics[width=9.5cm]{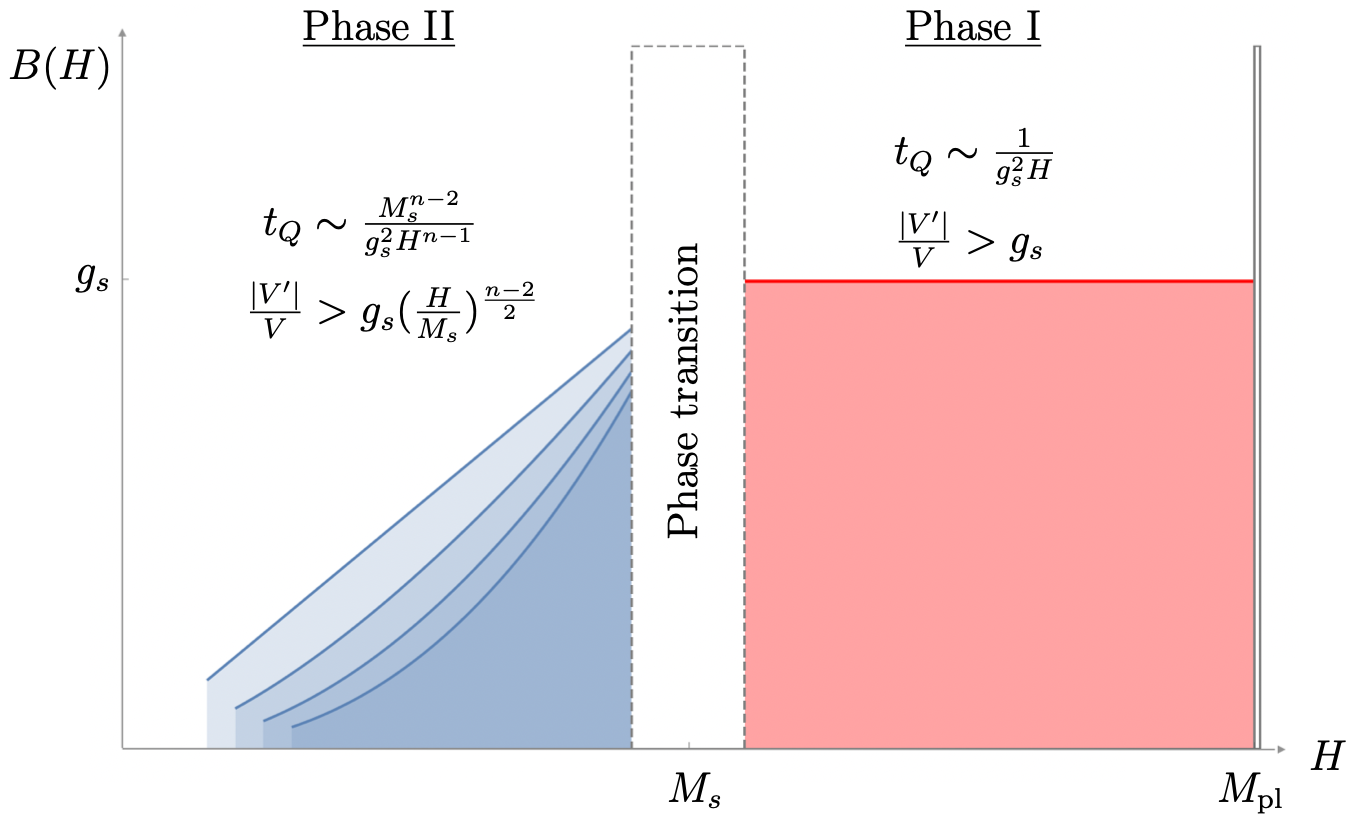}
  \caption{Schematic behavior of  lower bound $B(H)$ in the perturbative case with
     $M_s\ll M_{\rm pl}$ for dimensions $n=4,5,6,7$.  Shading corresponds to quantum breaking excluded region for  $|V'|/V$.}
  \label{fig:summary2}
\end{figure}

The induced quantum break
time $t_Q$ and the corresponding bound on $|V'|/V$, denoted as $B(H)$,
are shown in figure \ref{fig:summary2}. It is clear that the universal
quadratic scaling leads to a stronger bound for $|V'|/V$.  Note that
in these simplified figures  we are ignoring the 
$n$-dependent constant prefactors.

\noindent
Let us end this section with a couple of remarks.
We indeed found  the same bounds as  have been derived from
    the TCC, though now in different energy regimes. In phase II
    we find the weaker no eternal inflation bound, that is now 
     {\it the} bound in all quasi dS models resulting
    from string theory in phase II, i.e. with  $H<M_s$. Note that this
    seems to be  consistent with TCC, where
   the stronger dS swampland bound only applies asymptotically,
   whereas as demonstrated in \cite{Bedroya:2020rac} one can find effective
   potentials satisfying TCC but just marginally excluding eternal
   inflation. 
 
  In the quantum break time approach, the strong bound from the dS
   swampland conjecture is {\it the} bound in phase I, a regime that
   has not yet been investigated for dS vacua  in string theory/quantum
   gravity. We note that the asymptotic field limit in phase II 
   is sort of at the boundary. In fact, according to the Emergent String
   Conjecture \cite{Lee:2019wij}, an infinite distance limit in any quantum gravity
   theory is either an effective 
  decompactification limit or a limit in which a critical string
  becomes weakly coupled and  tensionless compared  to  the  Planck
  scale. In the latter case, it also has to be included in the thermal
  partition function. Therefore, both in the decompactification limit  
 and for the tensionless string case, in the limit   $\phi\to \infty$ 
   the string scale $M_s$ goes to zero and
   phase II disappears. This might be an explanation why for
   asymptotic field values one already encounters the stronger bound
   residing actually in phase I.

If indeed in the low $H$ phase 
 only the weaker constraint  $|V'|/ V\gtrsim c \,V^{n-2\over 4}$ was
 imposed by quantum gravity, it
would leave much more room for string theory realizations of
inflation and quintessence.  For instance,  in four space-time dimensions it leads to the maximal
number of e-foldings ${\cal N}_{\rm max}\sim M_s^2/H^2$.
On the other side, the stronger constraint in phase I
would forbid a sufficiently long period of inflation in this high temperature
phase of the universe leaving room for  the alternative
scenario of Agrawal, Gukov, Obied, Vafa  \cite{Agrawal:2020xek},  
utilizing precisely the presumed topological gravity theory valid in this regime.
Of course, all this is still very speculative and more research is
needed.

As observed recently in \cite{Bedroya:2020rmd}, up to
    logarithmic corrections, the 
 stronger bound in phase I   implies that the life-time $t_{\rm dec}$ of dS is shorter than the
scrambling/ther\-ma\-li\-za\-tion time\footnote{For another estimate on
  dS thermalization time see \cite{Danielsson:2003wb}.}  $t_{s}\sim t_Q$. Thus, one might suspect that 
our argument breaks down, as we were  initially assuming that
the system is in the thermal Bunch-Davies vacuum.
However, logically our arguments and conclusions are consistent
with \cite{Bedroya:2020rmd}. The logical chain can be summarized as
\eq{
   & {\rm dS\ is\ thermal\ } {\rm with}\  t_{\rm dec}> t_{s} \Longrightarrow
      {\rm QB\ takes\ place +censorship} 
\Rightarrow \\[0.2cm]
&\Rightarrow t_{\rm dec}<t_{Q}  \Longrightarrow {\rm not\
  thermal\ as\ } t_{\rm dec}< t_{s}\, ,
}
which is a contradiction to the initial assumption. Therefore, this
assumption is not correct and  the system decays before thermalization. 
But that is exactly what we intended to show, so our analysis can also
be seen as a consistency check.

\section{Conclusions}
\label{sec_concl}

In this paper we have developed a line of arguments that mutually
support the various assumptions made. We are not claiming
that all steps are bulletproof and beyond doubt, but we think that
we have at least uncovered potentially intricate relationships among
some of the very basic issues in string theory/quantum gravity both
from a bottom-up semiclassical approach and from a top-down
string theoretic approach.

\begin{figure}[htb]
  \centering
   \includegraphics[width=14cm]{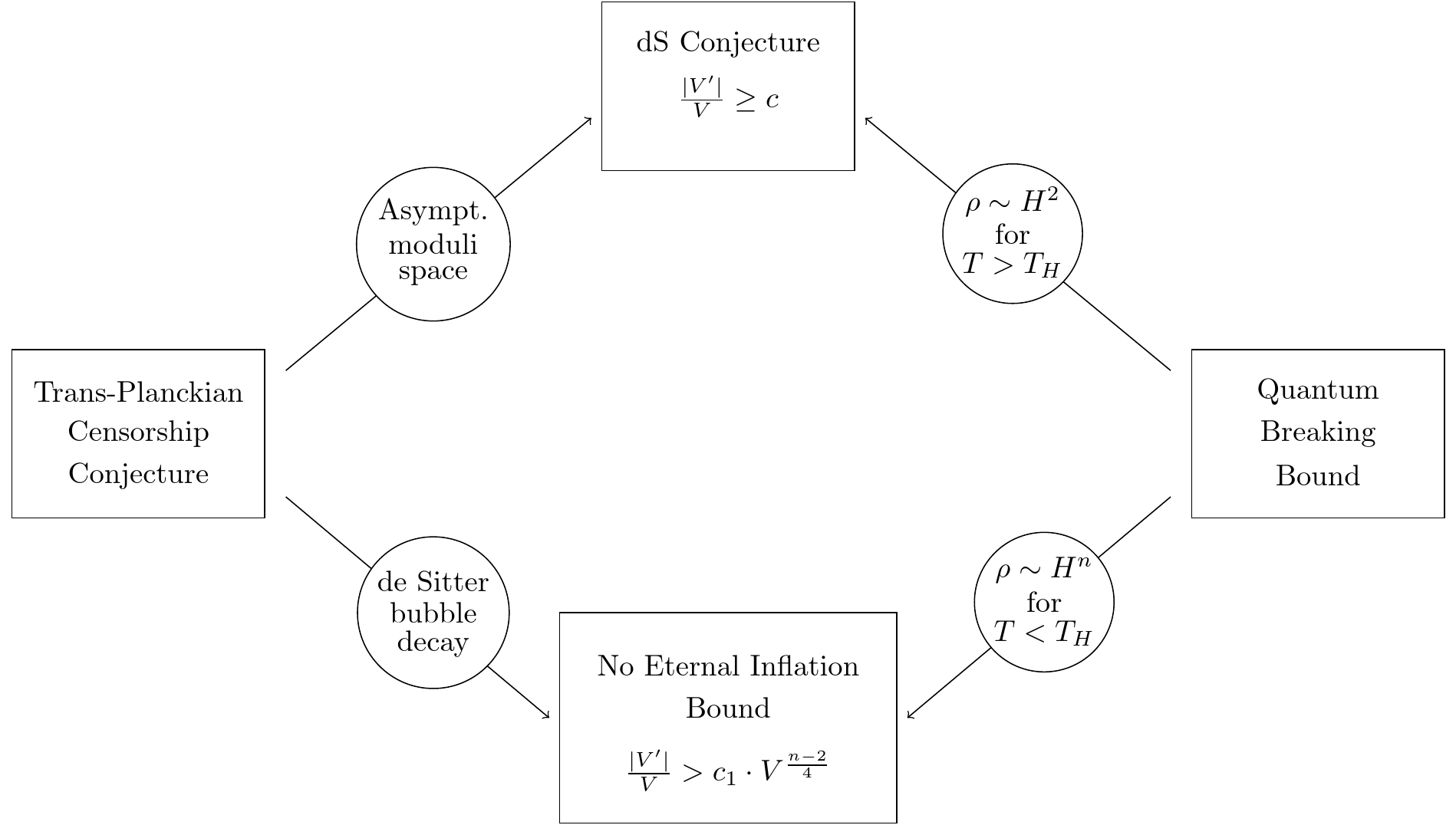}
  \caption{Schematic depiction of bounds for n-dimensional $V$ and their origin.}
  \label{fig:rev_overview}
\end{figure}

More concretely, as shown in  the (now complete) figure \ref{fig:rev_overview}, we have argued that the censorship of quantum breaking
does imply both the no eternal inflation bound and the (refined)
dS swampland conjecture in the low and high energy regimes,
respectively, under the following   assumptions:
\begin{itemize}
\item{Quantum breaking can be determined by taking the backreaction of
    the   semiclassical   quantum energy-momentum tensor in the static patch into account.}
\item{The essential scaling with the Gibbons-Hawking temperature can
    be deduced already from the analogous flat space energy-momentum tensor.
    The different IR behavior from modes with $\omega<m$ does only affect overall numerical prefactors.}
\item{All these features extend to string theory on dS spaces in a straightforward manner.}
 \item{In the high temperature phase of string theory,
     independently of the number of space time dimensions,  the 
     energy density universally scales as $\rho\sim T^2$ and at leading
     order the equation of motion  is still given by Einstein's equations of gravity.} 
\end{itemize}

\noindent
If true, the consequences of  censoring  quantum
breaking are very similar to those that can be inferred from 
the trans-Planckian censorship
conjecture, but now  implying the no eternal inflation bound   for low 
and the refined dS swampland conjecture for high Gibbons-Hawking temperatures.
It would be worthwhile to find a more direct relation between  the
TCC and quantum breaking and understand the appearance
of the $\log$-corrections also from the quantum breaking perspective.
Moreover, it would be interesting to check whether the quantum breaking approach can also imply
some of the other swampland conjectures that are interconnected with
TCC, such as the  swampland distance conjecture.
Another intriguing question is whether the string theoretic high $T$  quantum break
time could also be derived directly in the framework of decoherence presented
in the work of Dvali-Gomez-Zell \cite{Dvali:2017eba}. 

As a a further consequence of  our assumptions we are led to the
conjecture that eternal  Minkowski space is not a solution of quantum
gravity, either. This seems to be a bold conjecture that however
has already some support from string theory constructions \cite{Palti:2020qlc}.
Of course, it would be interesting to study this further and in
particular to see how for extended supersymmetry $N\ge 2$  this conclusion
could be avoided.

Finally, maybe the most exciting aspect of our analysis is the
appearance of the high energy regime of string theory. 
This also prominently appeared
in the recent proposal  \cite{Agrawal:2020xek} that the early
history of the universe is not described by inflation but by a topological gravity theory. 
It would be intriguing to identify the nature of this topological
theory and perform computations directly there, as for instance
confirming the quadratic high temperature scaling of the free energy.

%%%%%%%%%%%%%%%%%%%%%%%%%%%%%%%%%%%%%%%%%%%%%%%

\vspace{0.5cm}

\noindent
\subsubsection*{Acknowledgments}
We would like to thank Irene Valenzuela for helpful correspondence and
Rafa {\'A}lvarez-Garc{\'i}a, Max Brinkmann, Marco Scalisi
and Lorenz Schlechter for useful discussions and comments on the draft.
We also thank Lars Aalsma, Tommi Markkanen, Gary Shiu and Jan Pieter van
der Schaar  for enlightening critical discussions on a first version of this paper.

\vspace{1cm}

\clearpage

\appendix
\section{Computation of normalization constants}
\label{app_A}
In section \ref{sec_Markkanen_generalization} we mentioned that by imposing the commutation relations we can determine the normalization constants as \eqref{normalization_constant}.
Here we perform the corresponding calculations.
We start by writing down the commutation relations of $\Phi(r,\theta, \tau)$
\eq{
%\label{app_commrelPhi1}
[\Phi(r,\theta , \tau), &\dot{\Phi}(r',\theta', \tau)] = - \frac{i}{g^{\tau \tau}\sqrt{-g}}\, \delta(r-r')\, \delta(\theta- \theta') \, , \\
}
\eq{
\label{app_commrelPhi2}
[\Phi(r,\theta, \tau), &\Phi(r',\theta', \tau)] = [\dot{\Phi}(r,\theta, \tau), \dot{\Phi}(r',\theta', \tau)] = 0\,.
}
Furthermore, by requiring the usual commutation relations for the creation and annihilation operators
\eq{
%\label{app_commrela}
[\hat{a}_{L \lambda \omega}, \hat{a}^{\dagger}_{L' \lambda' \omega'}] =   \delta(\omega-\omega') \, \delta_{LL'}\, \delta_{\lambda \lambda'} \, , \\
[\hat{a}_{L \lambda \omega}, \hat{a}_{L' \lambda' \omega'}] = [\hat{a}^{\dagger}_{L \lambda \omega}, \hat{a}^{\dagger}_{L' \lambda' \omega'}] = 0
}
we get
\eq{
\label{inner_product}
-i \int d\Omega \int_{0}^{\frac{1}{H}} dr \, \frac{r^{n-2}}{1-H^2r^2} \, \Phi_1 \overleftrightarrow{\nabla}_{\tau} \, \Phi_{2}^{*} \overset{!}{=} \delta(\omega-\omega') \,\delta_{L L'}\, \delta_{\lambda \lambda'}   \,.  
} 
Due to the established orthogonality of the spherical harmonics this reduces to an equation purely for $f_{L \omega}(r)$
\eq{
\label{inner_product_R}
\int_{0}^{\frac{1}{H}} dr \, \frac{r^{n-2}}{1-H^2r^2}\,f_{L \omega}(r)\,f_{L \omega'}^{*}(r)  =\frac{1}{2\omega\, |N_{L \omega}|^2}\, \delta(\omega-\omega')  \,.
}
To solve this integral we may utilize the equations of motion \eqref{radialeom} to rewrite
\eq{
\label{identityeom}
&(\omega^2 - \omega' \,^2) \int_{0}^{\frac{1}{H}(1-\frac{\epsilon}{2})} dr \, \frac{r^{n-2}}{1-H^2r^2}\,f_{L \omega}(r)\,f_{L \omega'}^{*}(r) = \\
& r^{n-2}(1-H^2r^2)\Big(f_{L \omega}(r) \frac{d}{dr}f_{L \omega'}(r) - f_{L \omega'}(r) \frac{d}{dr}f_{L \omega}(r)  \Big)\bigg|_{r \rightarrow \frac{1}{H}(1-\frac{\epsilon}{2})}.
}
Additionally, the hypergeometric function has nice transformation properties. With formula (15.3.6) of \cite{mabramowitz64:handbook} we may recast $f_{L \omega}(r)$ given by \eqref{sol_radialeom} at
$r \sim \frac{1}{H}$ as
\eq{
\label{hypergeo_transform}
f_{L \omega}(r) \sim &\frac{\Gamma( \textstyle L + \frac{n-1}{2})\, \Gamma(\textstyle -\frac{i\omega}{H})}{\Gamma(\textstyle \frac{1}{2}(L - \frac{i\omega}{H} +\mu_{+}))\, \Gamma(\textstyle \frac{1}{2}(L - \frac{i\omega}{H} +\mu_{-}))}\,(1-H^2r^2)^{\frac{i\omega}{2H}} \, + \\
& \frac{\Gamma(\textstyle L + \frac{n-1}{2})\, \Gamma(\textstyle \frac{i\omega}{H})}{\Gamma(\textstyle \frac{1}{2}(L + \frac{i\omega}{H} +\mu_{+})) \, \Gamma(\textstyle \frac{1}{2}(L + \frac{i\omega}{H} +\mu_{-}))}\,(1-H^2r^2)^{-\frac{i\omega}{2H}}\,.
}
Using \eqref{identityeom} and \eqref{hypergeo_transform} we find \eqref{inner_product_R} to be
\eq{
\label{inner_product_R_2}
&I_{\epsilon} = \int_{0}^{\frac{1}{H}(1-\frac{\epsilon}{2})} dr \, \frac{r^{n-2}}{1-H^2r^2} \, f_{L \omega}(r) \, f_{L \omega}^{*}(r)  =\\ 
& \textstyle \frac{i}{H^{n-2}(\omega + \omega')}\Big[n^L_{\omega'\,}n^L_{\omega}\,e^{\frac{i}{2H}(\omega + \omega')\ln(\epsilon)}- n^L_{-\omega'}\,n^L_{-\omega}\,e^{-\frac{i}{2H}(\omega + \omega')\ln(\epsilon)}\Big] +\\ 
& \textstyle \frac{i}{H^{n-2} (\omega' - \omega)}\Big[n^L_{\omega'}\,n^L_{-\omega}\,e^{\frac{i}{2H}(\omega' - \omega)\ln(\epsilon)} - n^L_{-\omega'}\,n^L_{\omega}\,e^{-\frac{i}{2H}(\omega - \omega')\ln(\epsilon)}\Big],\,
}
with
\eq{
\label{n_constant}
n^L_{\omega} = \frac{\Gamma(\textstyle L + \frac{n-1}{2}) \, \Gamma(\textstyle \frac{i\omega}{H})}{\Gamma(\textstyle \frac{1}{2}(L + \frac{i\omega}{H} +\mu_{+}))\, \Gamma(\textstyle \frac{1}{2}(L + \frac{i\omega}{H} +\mu_{-}))}\,.
}  
By dropping the stronger oscillating terms we can write: 
\eq{
\label{inner_product_R_3}
I_{\epsilon} = \frac{2 \, |n^L_{\omega}|^2}{H^{n-2} \, (\omega' - \omega)} \sin \Big(\textstyle \frac{(\omega' - \omega)}{2H}\ln\big(\textstyle \frac{1}{\epsilon}\big)\Big) \xrightarrow{\epsilon \rightarrow 0} 2 \pi |n^L_{\omega}|^2 \delta(\omega - \omega'),
}
where we used $\lim\limits_{\alpha \to 0} \frac{\sin(\alpha x)}{x} \rightarrow \pi \delta(x)$. \\
By comparing \eqref{inner_product_R} and \eqref{inner_product_R_3} we can finally read off the normalization $|N_{L \omega}|^2$ as
\eq{
\label{app_normalization_constant}
|N_{L \omega}|^2 = \frac{H^{n-2}}{4 \pi \omega }\,
\frac{\big|\Gamma(\textstyle \frac{1}{2}(L + \frac{i\omega}{H}
  +\mu_{+}))\big|^2\; \big|\Gamma(\textstyle \frac{1}{2}(L + \frac{i\omega}{H} +\mu_{-}))\big|^2} 
{\big|\Gamma(\textstyle L + \frac{n-1}{2})\big|^2\;
  \big|\Gamma(\textstyle \frac{i\omega}{H})\big|^2}\,.
}

\section{Computation of individual contributions to $T_{00}$ and $T_{rr}$}
\label{app_B}

In this appendix we provide the general expressions for the constituents of the energy-momentum tensor components.
We see that to first order in ${O}(Hr)$ $\langle g^{rr}m^2 \Phi^2 \rangle$ and $- \langle g^{00}m^2 \Phi^2\rangle$ are identical:
\eq{
\langle g^{rr}m^2 \Phi^2 \rangle &= m^2 \langle \Phi^2 \rangle +{O}(Hr)^2 \, ,\\
- \langle g^{00}m^2 \Phi^2 \rangle &= m^2 \langle \Phi^2 \rangle +{O}(Hr)^2 \, .\\
}
$m^2 \langle \Phi^2 \rangle$ can be computed by using the $L=0$ mode expansion to first order in ${O}(Hr)$ \eqref{zero_mode_approx}:
\eq{
\label{exp_val_m}
 m^2 \langle \Phi^2 \rangle &= m^2 \Big \langle \Big | \int_0^{\infty}d\omega \, [\Phi_{0,\lambda = 0, \omega} \hat{a}_{0,\lambda = 0, \omega} + \text{H.C.}]\Big |^2 \Big \rangle\\
&= m^2 \int_0^{\infty}d\omega \, | \Phi_{0,\lambda = 0, \omega}|^2 \left[1 + 2 \langle \hat{n}_{0,\lambda = 0, \omega}\rangle \right] \\
&= m^2 \int_0^{\infty}d\omega \, \frac{H^{n-3} \sinh \left[\textstyle \frac{\pi \omega}{H} \right]}{4 \pi^2  \Gamma(\frac{n-1}{2})^2}\, |\Gamma(\textstyle \frac{1}{2}(\frac{i\omega}{H} +\mu_{+}))|^2 \,
|\Gamma(\textstyle \frac{1}{2}( \frac{i\omega}{H} +\mu_{-}))|^2 \\
& \hspace{5.75 cm} \times |Y_{0, \lambda = 0}|^2 \left[1 + \frac{2}{e^{2 \pi \omega/H} -1} \right] \\
&=  m^2 \int_0^{\infty}d\omega \, \frac{H^{n-3} \sinh \left[\textstyle \frac{\pi \omega}{H} \right]}{4 \pi^{\frac{n+3}{2}}  \Gamma(\textstyle \frac{n-1}{2})} \, |\Gamma(\textstyle\frac{1}{2}(\frac{i\omega}{H} +\mu_{+}))|^2 
\, |\Gamma(\textstyle \frac{1}{2}(\frac{i\omega}{H} +\mu_{-}))|^2 \\
& \hspace{9.15 cm} \times  \Theta\!\left({\textstyle \omega\over H}\right) \,.
%&=  m^2 \int_0^{\infty}d\omega \frac{1}{4 \pi^{\frac{n-1}{2}}  \Gamma(\frac{n-1}{2})} \bigg[\pi \omega \Gamma[\mu_{+}/2]^2 \Gamma[\mu_{-}/2]^2 + \frac{\pi \omega^3}{12H^2} \Gamma[\mu_{+}/2]^2 \Gamma[\mu_{-}/2]^2 \\
%&\quad \times \left(2\pi^2 -3 \psi^{(1)}(\mu_{-}/2) - 3 \psi^{(1)}(\mu_{+}/2)\right) +\frac{\pi \omega^5}{960H^4} \Gamma[\mu_{+}/2]^2 \Gamma[\mu_{-}/2]^2 \\
%&\bigg(8\pi^4 +30 (\psi^{(1)}(\mu_{-}/2) + \psi^{(1)}(\mu_{+}/2))^2 -40 \pi^2 \left(\psi^{(1)}(\mu_{-}/2) + \psi^{(1)}(\mu_{+}/2) \right) \\
%&+ 5 \left(\psi^{(3)}(\mu_{-}/2) + \psi^{(3)}(\mu_{+}/2) \right) + \mathcal O(\omega^7)\bigg) \bigg] \left[\frac{1}{2} + \frac{1}{e^{2 \pi \omega/H} -1} \right] 
}
%Obviously we cannot neglect any order in the Taylor series of the integrand in $\omega$, but what becomes apparent is the form of the expansion: $A_{o}(n,m)\omega^{o+1}/H^{o}$, where $o$ is the expansion order and $A_{o}(n,m)$ is a constant dependant on just the mass parameter m and the number of dimensions. This structure will be important for evaluating the Bose-Einstein integral.
Similarly the other contributions to $T_{00}$ and  $T_{rr}$ can be calculated,
first off the temporal derivative term:
\eq{
\label{exp_val_del0}
 \langle g^{00} (\del_0 \Phi)^2 \rangle &= - \langle (\del_0 \Phi)^2 \rangle +{O}(Hr)^2 \\
- \langle (\del_0 \Phi)^2 \rangle &=- \Big \langle \Big | \int_0^{\infty}d\omega[ -i\omega \Phi_{0,\lambda = 0, \omega} \hat{a}_{0,\lambda = 0, \omega} +\text{H.C.}]\Big |^2\Big \rangle\\
&= \int_0^{\infty}d\omega \, \omega^2 \, | \Phi_{0,\lambda = 0, \omega}|^2 \, \left[1 + 2 \langle \hat{n}_{0,\lambda = 0, \omega}\rangle \right] \\
%&= \int_0^{\infty}d\omega \frac{H^{n-3} \omega^2 \sinh \left[\textstyle \frac{\pi \omega}{H} \right]}{4 \pi^2  \Gamma(\textstyle \frac{n-1}{2})^2} |\Gamma(\textstyle \frac{1}{2}(\frac{i\omega}{H} +\mu_{+}))|^2 
% |\Gamma(\textstyle \frac{1}{2}(\frac{i\omega}{H} +\mu_{-}))|^2\\
%& \hspace{5.65 cm} \times |Y_{0, \lambda = 0}|^2 \left[1 + \frac{2}{e^{2 \pi \omega/H} -1} \right] \\
&=  \int_0^{\infty}d\omega \, \frac{H^{n-3} \, \omega^2 \sinh \left[\textstyle \frac{\pi \omega}{H} \right]}{4 \pi^{\frac{n+3}{2}} \, \Gamma(\textstyle \frac{n-1}{2})} |\Gamma(\textstyle \frac{1}{2}(\frac{i\omega}{H} +\mu_{+}))|^2 \, |\Gamma(\textstyle \frac{1}{2}(\frac{i\omega}{H} +\mu_{-}))|^2 \\
& \hspace{9 cm} \times \Theta\!\left({\textstyle \omega\over H}\right).
%&=  \int_0^{\infty}d\omega \frac{\omega^2}{4 \pi^{\frac{n-1}{2}}  \Gamma(\frac{n-1}{2})} \bigg[\pi \omega \Gamma[\mu_{+}/2]^2 \Gamma[\mu_{-}/2]^2 + \frac{\pi \omega^3}{12H^2} \Gamma[\mu_{+}/2]^2 \Gamma[\mu_{-}/2]^2 \\
%&\quad \times \left(2\pi^2 -3 \psi^{(1)}(\mu_{-}/2) - 3 \psi^{(1)}(\mu_{+}/2)\right) +\frac{\pi \omega^5}{960H^4} \Gamma[\mu_{+}/2]^2 \Gamma[\mu_{-}/2]^2 \\
%&\bigg(8\pi^4 +30 (\psi^{(1)}(\mu_{-}/2) + \psi^{(1)}(\mu_{+}/2))^2 -40 \pi^2 \left(\psi^{(1)}(\mu_{-}/2) + \psi^{(1)}(\mu_{+}/2) \right) \\
%&+ 5 \left(\psi^{(3)}(\mu_{-}/2) + \psi^{(3)}(\mu_{+}/2) \right) + \mathcal O(\omega^7)\bigg) \bigg] \left[\frac{1}{2} + \frac{1}{e^{2 \pi \omega/H} -1} \right] 
}
The radial derivative term is then given by:
\eq{
\label{exp_val_delr}
\langle g^{rr} (\del_r \Phi)^2 \rangle &= \langle (\del_r \Phi)^2 \rangle +{O}(Hr)^2 \\
\langle (\del_r \Phi)^2 \rangle &=  \Big \langle \Big | \del_r \sum_{\lambda} \int_0^{\infty}d\omega \, [\Phi_{1,\lambda, \omega} \hat{a}_{1,\lambda, \omega} +\text{H.C.}] \Big |^2\Big \rangle \\
%&= \int_0^{\infty}d\omega \frac{H^{n-1}\sinh \left[\textstyle \frac{\pi \omega}{H} \right]}{4 \pi^2  \Gamma(\textstyle \frac{n+1}{2})^2} |\Gamma(\textstyle \frac{1}{2}(1 +\frac{i\omega}{H} +\mu_{+}))|^2 \\
%&\hspace{2.0 cm} \times |\Gamma(\textstyle \frac{1}{2}(1 +\frac{i\omega}{H} +\mu_{-}))|^2 \sum_{\lambda}|Y_{1, \lambda}|^2 \left[1 + \frac{2}{e^{2 \pi \omega/H} -1} \right] \\
&= \int_0^{\infty}d\omega \, \frac{(n-1)H^{n-1} \, \Gamma(\textstyle \frac{n-1}{2})\sinh \left[\textstyle \frac{\pi \omega}{H} \right]}{4 \pi^{\frac{n+3}{2}} \, \Gamma(\textstyle \frac{n+1}{2})^2} \, |\Gamma(\textstyle \frac{1}{2}(1 +\frac{i\omega}{H} +\mu_{+}))|^2 \\
& \hspace{5.5 cm} \times |\Gamma(\textstyle \frac{1}{2}(1 +\frac{i\omega}{H} +\mu_{-}))|^2 
\, \Theta\!\left({\textstyle \omega \over H}\right).
%&=  \int_0^{\infty}d\omega \frac{n-3}{(n-1)^3 \pi^2  \Gamma(\frac{n-1}{2})} \bigg[\pi H^2 \omega \Gamma[1+\mu_{+}/2]^2 \Gamma[1+\mu_{-}/2]^2 \\
%&\quad+ \frac{\pi \omega^3}{12} \Gamma[(1+\mu_{+})/2]^2 \Gamma[(1+\mu_{-})/2]^2 \bigg(2\pi^2 -3 \psi^{(1)}((1+\mu_{-})/2) \\
%&\quad- 3 \psi^{(1)}((1+\mu_{+})/2)\bigg) +\frac{\pi \omega^5}{960H^2} \Gamma[(1+\mu_{+})/2]^2 \Gamma[(1+\mu_{-})/2]^2 \\
%&\quad \times \bigg(8\pi^4 +30 (\psi^{(1)}((1+\mu_{-})/2) + \psi^{(1)}((1+\mu_{+})/2))^2 \\
%&\quad -40 \pi^2 \left(\psi^{(1)}((1+\mu_{-})/2) + \psi^{(1)}((1+\mu_{+})/2) \right) \\
%&\quad + 5 \left(\psi^{(3)}((1+\mu_{-})/2) + \psi^{(3)}((1+\mu_{+})/2) \right) + \mathcal O(\omega^7)\bigg) \bigg] \\
%& \quad \times \left[\frac{1}{2} + \frac{1}{e^{2 \pi \omega/H} -1} \right] 
}
All of the expressions above are evaluated to first order in ${O}(Hr)$, since we are interested
in the energy density and pressure for an observer far away from the
horizon.

\section{Limit $m\gg H$ in 4D}
\label{app_C}

In this appendix we want to prove that the equation of state parameter
for the 4D massive scalar field in the limit  $m\gg H$ is
$\omega=-2/3$.
First we observe that in this regime we can approximate
\eq{
  \mu_\pm={3\over 2}\pm {1\over 2}\sqrt{1-4{m^2\over H^2}}\approx
  {3\over 2}\pm i{m\over H}\,.
}  
Next we  employ  the  asymptotic formula for the gamma function \cite{mabramowitz64:handbook}
\eq{
\label{asymptotic_gamma}
\lim\limits_{|y| \to \infty} (2\pi)^{-\frac{1}{2}}\,\big|
\Gamma(x+iy)\big|\, e^{\frac{1}{2}\pi |y|}\,|y|^{\frac{1}{2}-x} = 1\,
}
to derive for the combination of $\Gamma$-functions appearing
in the general expressions \eqref{rhofinal4D} and \eqref{pfinal4D} for $\rho$ 
and $p$
\eq{
\big|\Gamma\big({\textstyle {3\over 4}+{i\omega\over 2H}(\omega+m) }\big)\big|^2
                         &\big|\Gamma\big({\textstyle {3\over 4}+{i\omega\over
                             2H}(\omega-m)}\big)\big|^2\\[0.1cm]
                         &=\begin{cases}
\frac{2\pi^2}{H} \,\sqrt{\omega^2-m^2}\,
e^{-\frac{\pi \omega}{H}}, & {\rm for\ }\omega>m\\[0.3cm]
\frac{2\pi^2}{H}  \,\sqrt{m^2-\omega^2} \,
e^{-\frac{\pi m}{H}}, &  {\rm for\ }\omega<m\,
\end{cases}
}
and
\eq{
\big|\Gamma\big({\textstyle {5\over 4}+{i\omega\over 2H}(\omega+m) }\big)\big|^2
                         &\big|\Gamma\big({\textstyle {5\over 4}+{i\omega\over
                             2H}(\omega-m)}\big)\big|^2\\[0.1cm]
                         &=\begin{cases}
\frac{\pi^2}{2H^3} \,\sqrt{\omega^2-m^2}^3\,
e^{-\frac{\pi \omega}{H}}, & {\rm for\ }\omega>m\\[0.3cm]
\frac{\pi^2}{2H^3}  \,\sqrt{m^2-\omega^2}^3 \,
e^{-\frac{\pi m}{H}}, &  {\rm for\ }\omega<m\,.
\end{cases}
}
Looking at the exponentials, it is clear that the main contribution comes
from the integration region $\omega<m$, i.e. the one that is not
present in the thermal flat space relation \eqref{rhofreegas}.
Therefore, the main contributions to $\rho$ and $p$ can be simplified
as
\eq{
     \rho\approx {1\over 3\pi^2}\int_0^m d\omega\, m^2\,
     \sqrt{m^2-\omega^2} \, e^{-\frac{\pi m}{H}}
      {\sinh({\pi \omega\over H})\over e^{2\pi\omega/H}-1} 
 }    
 and
 \eq{
     p\approx  -{2\over 9\pi^2}\int_0^m d\omega \,m^2\,
     \sqrt{m^2-\omega^2} \, e^{-\frac{\pi m}{H}}
      {\sinh({\pi \omega\over H})\over e^{2\pi\omega/H}-1} 
 }    
which apparently satisfy $p=-{2\over 3} \rho$.

\clearpage

\bibliography{references}  
\bibliographystyle{utphys}

%%%%%%%%%%%%%%%%%%%%%%%%%%%%%%%%%%%%%%%%%%%%%%%
%%%%%%%%%%%%%%%%%%%%%%%%%%%%%%%%%%%%%%%%%%%%%%%
%%%%%%%%%%%%%%%%%%%%%%%%%%%%%%%%%%%%%%%%%%%%%%%
%%%%%%%%%%%%%%%%%%%%%%%%%%%%%%%%%%%%%%%%%%%%%%%

\end{document}